\documentclass[a4paper,twocolumn,11pt,unpublished,]{quantumarticle}
\pdfoutput=1
\usepackage[utf8]{inputenc}
\usepackage[english]{babel}
\usepackage[T1]{fontenc}
\usepackage{amsmath}
\usepackage{hyperref}
\usepackage{tikz}
\usepackage{lipsum}
\usepackage{adjustbox} 
\usepackage[showframe=false]{geometry}
\usepackage{changepage}
\usepackage{placeins}
\usepackage{graphicx} 
\usepackage{caption} 
\usepackage{amssymb,amsfonts,dsfont}
\usepackage{comment}
\usepackage{braket}
\usepackage{tablefootnote}
\usetikzlibrary{quantikz2}

\begin{document}

\title{Optimised T counts and active volume estimates for high- and low- level arithmetic subroutines}

\author{Sam Heavey}
\email{sheavey@psiquantum.com}
\affiliation{Imperial College London, London}

\date{27/04/2025}
    
\maketitle

\begin{abstract}
Surface code based quantum computers show great promise for fault-tolerant quantum computing, but most architectures needlessly increase the spacetime volume of a computation due to qubits sitting idly during a computation. Active volume architectures, with long-range connectivity, aim to remove idle spacetime volume leaving only the spacetime volume that logically contributes to a computation. In this work we optimise and derive the active volumes for several industry-leading low- and high-level arithmetic subroutines and achieve significant T-count reductions. We discuss a simple method for estimating and optimising active volumes using orientated ZX diagrams. We also demonstrate that circuit structure, beyond gate counts alone, impacts the active volume of a subroutine when state injection is used and therefore should be taken into consideration when designing circuits.
\end{abstract}

\section{Introduction and background} 
\label{sec:introduction}
Quantum computers have the capacity to exponentially reduce the computation time of classically difficult problems.
However, useful quantum computations require error correction schemes \cite{gidney2021factor, chakrabarti2021threshold, von2021quantum} to prevent environment-induced errors from corrupting results. 
One of the most promising error correction schemes is that of surface codes \cite{dennis2002topological, kitaev2003fault, litinski2019surface}. In the context of surface codes, a logical qubit is represented by a surface code patch, which is a $d \times d$ array of physical qubits. 
Therefore, $d$, the code distance parameterises the cost of a logical qubit. 
To execute a computation, individual surface code patches must be complied in some architecture.
In this work we discuss two types of general-purpose surface code architectures, baseline architectures and the active volume architecture. The distinction between the two architectures is captured in \cite{litinski2022active}, which also introduces the active volume architecture. 
\par \textbf{Baseline architectures.} Baseline architectures are the most prominent in literature and describe a 2-D array of locally connected surface code patches \cite{litinski2019surface, fowler2018low, chamberland2022universal, bombin2021interleaving, chamberland2022building}.
When using the Clifford+T universal gate set, the cost of a baseline computation scales with the circuit volume $n_Q \times n_T $, where $n_Q$ is the total number of qubits and $n_T$ is the total number of T gates \cite{litinski2022active, litinski2023compute}.
T gates and other non-Clifford gates (e.g. Toffoli) require the consumption of magic states and are vastly more expensive than Clifford gates (e.g. Hadamard, S gate, CNOT) \cite{chen2025efficientmagicstatecultivation, bravyi2005universal, bravyi2012magic}. For this reason, Clifford counts for algorithms are typically ignored in literature.
\par Due to local connectivity, a large portion of the circuit volume may consist of 'idle' volume, which corresponds to logical qubits that sit idly during the computation. For example, a 2000 qubit computation using Toffoli gates which only acts on a small subset of the total qubits each logical cycle can have up to $99\%$ idle volume \cite{litinski2022active}.
This idle volume adds to the spacetime cost of a computation without contributing to the its logic. Note that to eliminate idle volume, we look to complete operations in parallel, which  we requires the creation of bridge qubits \cite{litinski2022active}. This means if we eliminate $x\%$ idle volume, typically we do not decrease the computation cost by $x\%$. However, often this trade off is not proportional and we see a significant reduction in overall spacetime volume, which makes parallelisation advantageous.
\par \textbf{Active volume architecture.} Active volume (AV) refers to non-idle spacetime volume and has units of logical blocks. 
AV corresponds to the volume of qubits involved in logical operations and therefore has contributions from both Clifford and non-Clifford gates.
Importantly, AV does not scale with the number of qubits like the circuit volume in baseline computers. Instead, the AV of a computation is calculated by summing the AV of its constituent operations. AV computers are effective at enabling the parallelisation of operations, eliminating idle volume, and leaving primarily the active volume to contribute to the overall spacetime cost. 
This reduction in spacetime volume means that compared to a baseline computer with the same number of physical qubits (space) an active volume computation can be computed more quickly (time). 
For example, Litinski found the 2048-bit factoring algorithm has an active volume of $8.7 \times 10^{11}$ and a baseline circuit volume of $3.8 \time 10^{13}$. Running the algorithm in each architecture reveals a $53$ times reduction in the computation time for active volume computation compared to the baseline computation \cite{litinski2022active}. 
Alternatively, we can use fewer physical qubits to complete a computation in the same time as a baseline architecture.
The architecture achieves this cost reduction by allowing for non-local connectivity between logical qubits. 
An active volume computer with $n$ logical qubits allocates half of those logical qubits as workspace modules and half as memory modules.
Each workspace module executes a logical block after $d$ code cycles. The number of workspace modules therefore determines the speed of a computation.
Memory modules are used to store magic states and data qubits (the input qubits to a circuit) that are not involved in a round of operations being completed in the workspace modules. 
For this reason, the memory capacity of a computer determines which computations it is able to complete.
Memory modules are also used to move qubits in storage into ideal positions for the next set of operations. 
Note, these workspace modules can be used as memory modules (and vice versa) during a computation if needed \cite{litinski2023compute}. 
The long range connectivity required for the active volume architecture is best suited for photonic computing, but is possible for matter-based computers, such as those using superconducting or trapped-ion qubits \cite{litinski2022active, litinski2023compute}. 
\par \textbf{Active volume computing: Previous work and resource estimates.} Quantum algorithms are composed of smaller operations called subroutines (e.g. low-level quantum arithmetic such as addition or subtraction) \cite{haner2018arithmetic, gidney2021factor, litinski2023compute, bhaskar2015, wang2020tran, sqrt_circuit, haner2018polyfit, edgard2019mult, thapliyal2017divide}. 
These optimised subroutines are used modularly such that they are inserted into algorithms when required. The total active volume of a circuit is simply given by the sum of the active volumes of each subroutine.
Consequently, a catalogue of optimised subroutines and their active volumes would allow for easy resource calculations and help guide efficient algorithm design.
Litinski began this process in \cite{litinski2022active, litinski2023compute}, which provides the active volume costs for many elementary gates (e.g. Hadarmard, CNOT, Toffoli) and some basic arithmetic (e.g. addition, controlled addition). Litinski also provides the distillation costs/estimates for of T, CCZ, and Y magic states. 
But how is the active volume of an operation calculated?
\par \textbf{ZX diagrams.}
ZX calculus graphically represents linear maps between qubits in a topologically free environment and therefore can be seen as a generalisation of quantum circuit notation \cite{Coecke2011, zxcalculus2020}. 
A ZX diagram consists of Z spiders, X spiders, and wires that connect them, where Z (X) spiders represent operations constructed from Pauli Z (X) eigenstates.
The number of wires entering (leaving) denote the number of qubits in the bras (kets). For example, a Z spider with one wire entering from the left and two wires leaving from the right represents $ \ket{00} \bra{0} + \ket{11} \bra{1}$.
Furthermore spiders can have a phase, which is applied to the $\ket{1}$ term for Z spiders and the $\ket{-}$ term for X spiders.
ZX calculus is topologically free because all meaningful information is represented by spider-spider connections such that spider positions do not matter.
This topological freedom combined with the rules and identities of ZX calculus make it very useful for circuit optimisation.

\begin{figure}[t]
  \centering
  \includegraphics[scale=0.47]{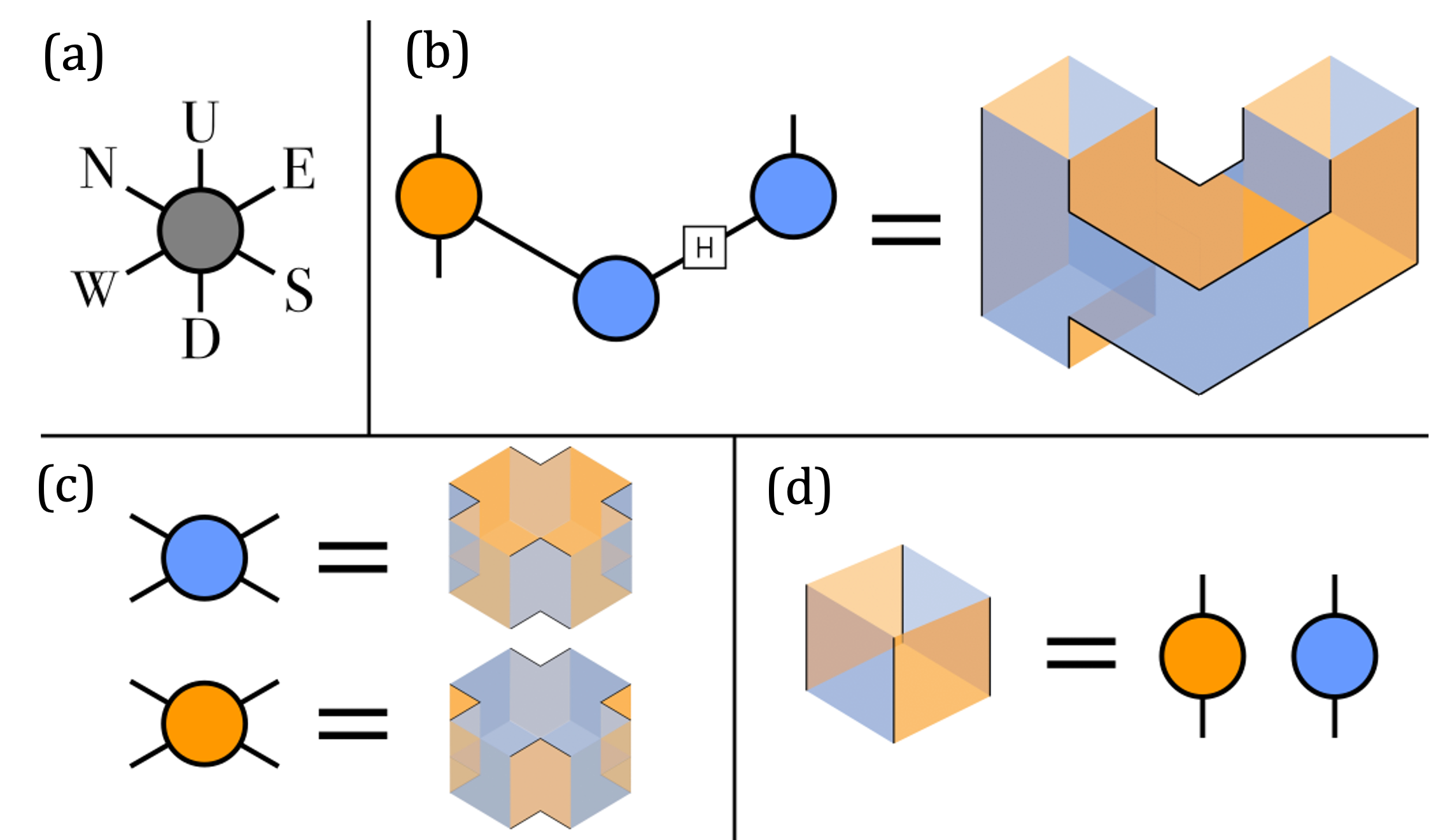} 
  \caption{(a) The locations of each port on a orientated spider. (b) An orientated ZX diagram and its surface code spacetime block equivalent. This example contains a Hadamard operation and spiders with 3 or 2 connected ports. The orange (blue) spacetime block boundaries correspond to surface code Z (X) boundaries. (c) Orientated ZX spiders with 4 connections and their spacetime equivalent. (d) The conventional orientation \cite{litinski2022active} of an idle spacetime block for both blue and orange spiders.}
  \label{fig:ZX_to_spacetime}
\end{figure}

\par \textbf{Orientated ZX diagrams.} Unlike ZX diagrams, orientated ZX diagrams (OZX) have an awareness of spatial orientation only allowing connections to six equally spaced points on the circumference of a spider. OZX spiders are restricted to having a phase of 0, such that these six points correspond to the U, D, N, S, E, W ports of 3D-spacetime blocks \cite{litinski2022active}. 
Connections between spiders in OZX diagrams obey the following connectivity rules \cite{litinski2022active}: All spiders can hold a maximum of four connections and must be either be U, N, or E orientated. A U orientated spider has no connections to the U or D ports, similarly an N orientated spider has no connections to the N or S ports and a E orientated spider has no E or W connections. 
Two spiders of the same type (Z or X) that share a direct connection must have the same orientation, whereas spiders of different types must have different orientations. 
However, this rule flips if the connection between the spiders is Hadamarded such that two spiders that are of different (same) types and share a Hadamarded connection must have the same (different) orientation.
For spiders with input/output states, the D port serves as the input port, and the U port serves as the output port.
By convention we orientate idle spacetime blocks such that the NS ports are associated with the Z boundaries of the surface code patch, while the EW ports are associated with X boundaries, see Fig.~\ref{fig:ZX_to_spacetime}. 
The idle block orientation convention enforces Z (X) spiders with an input or output to be E (N) oriented, so that the execution of a logical block network automatically produces correctly orientated idle qubits.
\par Strictly speaking for perfect one-to-one correspondence between orientated ZX diagrams and spacetime/logical blocks only port connections between ports facing the same direction are allowed (e.g. N to N or U to U) \cite{litinski2022active}. However, for orientated ZX diagrams this rule can be relaxed, in cases where symmetry allows (no loops/cycles connected using one axis that contain an odd number of spiders), such that ports can be connected to ports along the same axis (e.g. N to S or U to D). This lifting of restrictions is simply to improve readability of the diagram and has no effect on spider/block counts. 
\par \textbf{Overview.}
In Sections~\ref{sec:low_level_arithmetic} and \ref{sec:high_level_arithmetic} we provide active volume cost formulas and discuss baseline line modifications that reduce the T count for low- and high-level arithmetic, respectively. A summary table displaying the activate volume of all subroutines discussed in this paper as well as a table displaying their baseline costs can be found in section~\ref{sec:table_summaries} of the appendix. For more details on the active volume calculations please refer to sections~\ref{sec:av_details} and \ref{sec:derivations} of the appendix.
In the following Section~\ref{sec:method} we explain our optimisation methodology and the process for determining the active volume of a circuit.

\section{Active volume optimisation assumptions}
\label{sec:method}
The high-level arithmetic subroutines targeted in this paper are themselves constructed from low-level arithmetic such as multiplication and addition \cite{haner2018polyfit, wang2020tran, sqrt_circuit}. Therefore, to create optimised high-level arithmetic subroutines we focus on optimising their low-level arithmetic components.

\par In this paper we make the following assumption: circuit designs with the lowest T count and number of qubits, will also produce subroutines with the smallest active volume. This assumption spares us the need to calculate the active volumes of various designs as we only convert the cheapest baseline implementation. Although all gates contribute to the active volume of a subroutine T/Toffoli gates are by far the most expensive as they require the distillation and storage of magic states and reactive measurements, this is explained later in Section~\ref{sec:depth}. Furthermore, most of the baseline differences between adder-based subroutine implementations are due to the choice of adder from which they are constructed. Inspection of the various adder circuits used by this paper and others we note that each has a similar number of CNOT gates \cite{Gidney2018adder, thapliyal2016addsub, edgard2019mult, takahashi2009adder}. This means that the active volume contributed by CNOTs across various design should be roughly equivalent. This allows us to focus on reducing the commonly reported T and qubit counts. Finally, although we compare baseline T counts, as they are commonly reported in the literature, we will execute Toffoli and Temporary AND gates with CCZ magic states instead of T states, as they can be cheaper \cite{litinski2022active}. This means our active volume calculations will include the distillation cost of CCZ states, $C_{\ket{CCZ}}$, not T states. The distillation cost of a CCZ state is $\approx 35$ blocks \cite{litinski2022active, litinski2023compute}.

\par \textbf{Baseline optimisation.} First, we reduce the baseline cost of these operations such that their T count is minimised. We do this mainly by utilising the temporary AND gate introduced by Gidney, which can be used to halve the T count of certain operations, i.e. addition \cite{Gidney2018adder}. We construct the required low-level arithmetic subroutines using this gate and or modifications to the ripple carry adders presented by \cite{Gidney2018adder, litinski2022active}. Only Toffoli and temporary AND gates in CNOT-Toffoli-temporary AND circuits contribute to the T count, each requiring 7 T gates \cite{jones2013}. Therefore, the total T count of a subroutine is found by summing the number of Toffoli and Temporary AND gates and then multiplying by $7$.  Once we obtain circuits for all the low-level components, we proceed with their active volume calculation.  

\begin{figure*}[t]
\centering
\includegraphics[scale=0.9]{ 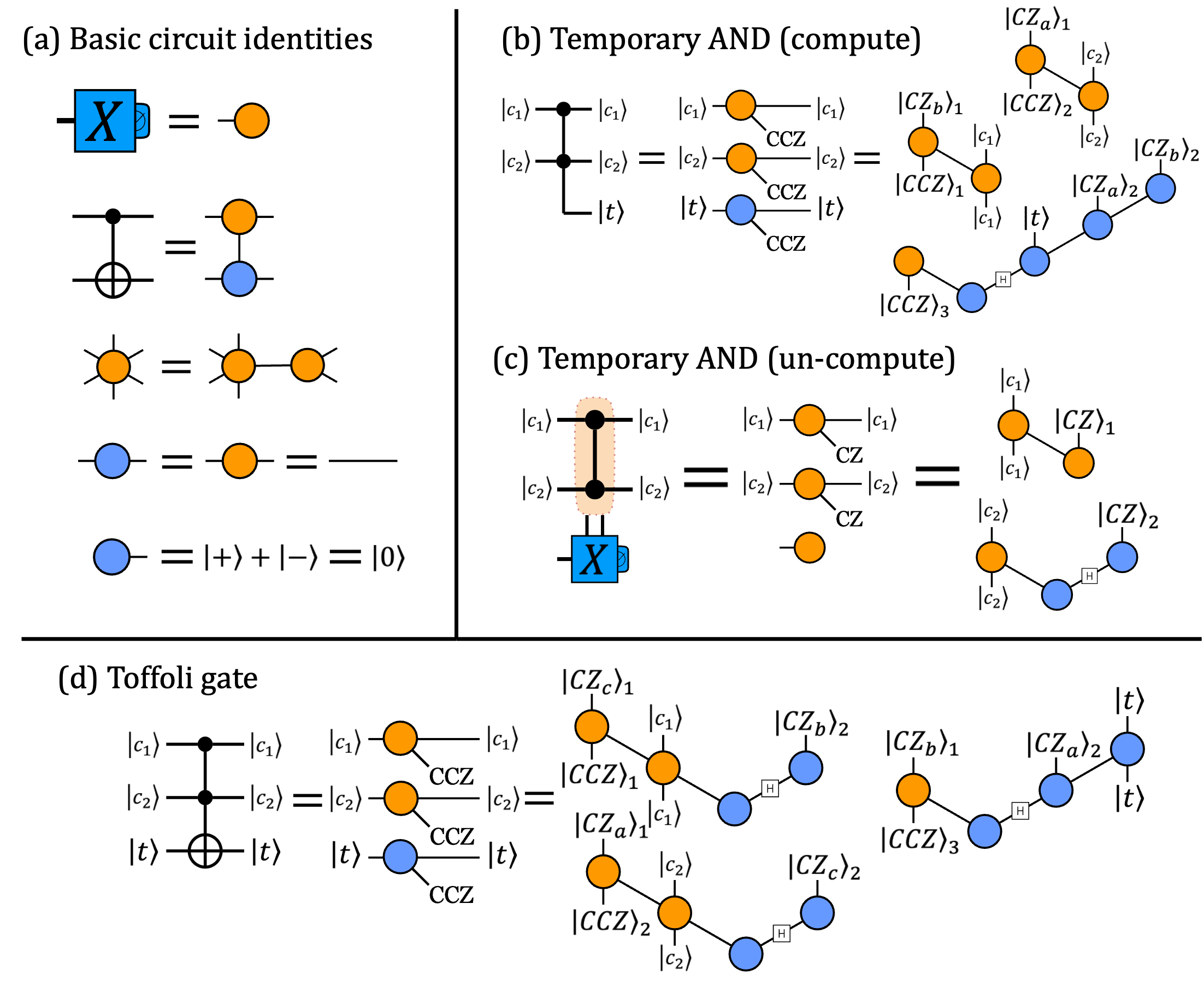} 
\caption{(a) Basic identities between circuit notation and ZX calculus. From top to bottom we have, a measurement in the X basis, a CNOT gate, an example of spider splitting/merging, the ZX equivalent of the identity matrix, and initialisation of the $\ket{0}$ state. (b) The first part of a temporary AND gate, it has a depth of 1 and costs 9 blocks. (c) the last part of a temporary AND gate, it has depth 1 and costs 5 blocks. (d) A Toffoli gate has depth 1 and a block cost of 12. To implement (b) and (d), the three qubits of a CCZ state are measured. Depending on the measurement outcomes we may require up to three corrective CZ gates, which are associated with the CZ states labeled by $a$, $b$, and $c$. The CZ gates are implemented or discarded reactively by measuring each CZ state's qubits, 1 and 2, in the appropriate basis \cite{litinski2022active}. Note, a temporary AND gate costs the same as a Toffoli gate when the last part, (c), commutes with all operations between it and the start (b), i.e. depth 1 and 12 blocks \cite{litinski2022active}.}
\label{fig:ZX_identities}
\end{figure*}

\subsection{Calculating the active volume}
\label{sec:active_volume_calculation}
The process of converting a baseline operation into an active volume operation can be partitioned into the following steps: 
\begin{enumerate} 
\item Circuit diagram
\item ZX diagram
\item Orientated ZX diagram
\end{enumerate}

\textit{Step 1.} Obtain the circuit diagram that performs the desired subroutine. In this paper, we work with circuits composed of CNOT and Toffoli/Temporary AND gates. However, \cite{litinski2022active} also provides tools for finding the active volume of circuits written with other gates/ gate sets, e.g., a sequence of Pauli product rotations and measurements \cite{litinski2019surface}. Note, that NOT (X) gates and Z gates are ignored as these can be tracked classically in the Pauli frame \cite{pauli_frame2012}. This means inverted controls cost the same as normal controls.

\par \textit{Step 2.} To convert the circuit into a ZX diagram we use the identities provided in Fig.~\ref{fig:ZX_identities}. Additional identities for circuits written in other forms can be found in \cite{litinski2022active, zxcalculus2020}. After constructing the ZX diagram we can create a compressed ZX diagram by using ZX calculus to simplify the circuit into as few spiders as possible. 

\par \textit{Step 3.} In this step we convert the compressed ZX diagram into an OZX diagram. We do this by expanding spiders with over 4 legs into an array of connecting spiders which we call 'islands'. The island must be large enough such that each spider in its body has less then 4 legs and satisfies the OZX connectivity rules to other spiders/islands. Finally, after constructing a valid OZX diagram, we simply count the number of spiders and add the distillation cost of any magic state inputs to obtain the active volume. Each spider corresponds to one block \cite{litinski2022active}.

\subsection{Reaction depth}
\label{sec:depth}
Non-Clifford gates, such as Toffoli and temporary AND gates, require the measurement of magic states \cite{bravyi2005universal, bravyi2012magic}. Depending on the measurement outcomes, we may require corrective operations, such as CZ gates. These corrections involve the classical processing of the results of the first measurement followed by the appropriate choice of reactive measurements. The duration of this process is called the reaction time, and it means that non-Clifford gates have a reaction depth. A reaction depth of $1$ corresponds to waiting $1$ reaction time before learning whether a correction is needed. The gate cannot be fully implemented until this time as passed. Each unit of reaction depth is typically assumed to be equal to $10 \, \mu s$ \cite{litinski2023compute}. The reaction depth of an operation therefore restricts the speed of a computation.
\par Logical operations on disjoint groups of qubits (or even joint groups if bridge qubits are utilised \cite{litinski2022active}) can be performed in parallel, we call these groups layers \cite{litinski2022active, litinski2023compute}. Each layer has a maximum reaction depth that is ‘seen’ by one or more of the qubits in that layer. This maximum depth acts as the reaction depth for the entire layer. However, the number of operations that can be completed in parallel depends on the number of workspace modules available. Therefore, we cannot determine the number of layers or their reaction depth until we know the size of the quantum computer being used. Due to this constraint, we calculate the upper bound of a circuit’s reaction depth by summing the total depth of its components. 
\par The Toffoli, temporary AND compute, and temporary AND un-compute gates—the non-Clifford gates relevant to this research—each have a reaction depth of 1. However, it is important to note that when the un-compute part of a temporary AND commutes with all operations between it and the compute part, the temporary AND pair costs the same as a Toffoli and therefore has a reaction depth of 1 \cite{litinski2022active}.

\section{Low-level arithmetic subroutines}
\label{sec:low_level_arithmetic}
In this section we create efficient low-level arithmetic subroutines and determine their active volumes. The subroutines presented here serve as component pieces for the high-level arithmetic subroutines in Section~\ref{sec:high_level_arithmetic} and have been verified by simulation using the Qiskit python package.

\subsection{Increment \& controlled increment} 
\label{sec:incrementors}
\begin{figure*}[t]
    \centering
    \includegraphics[scale=0.95]{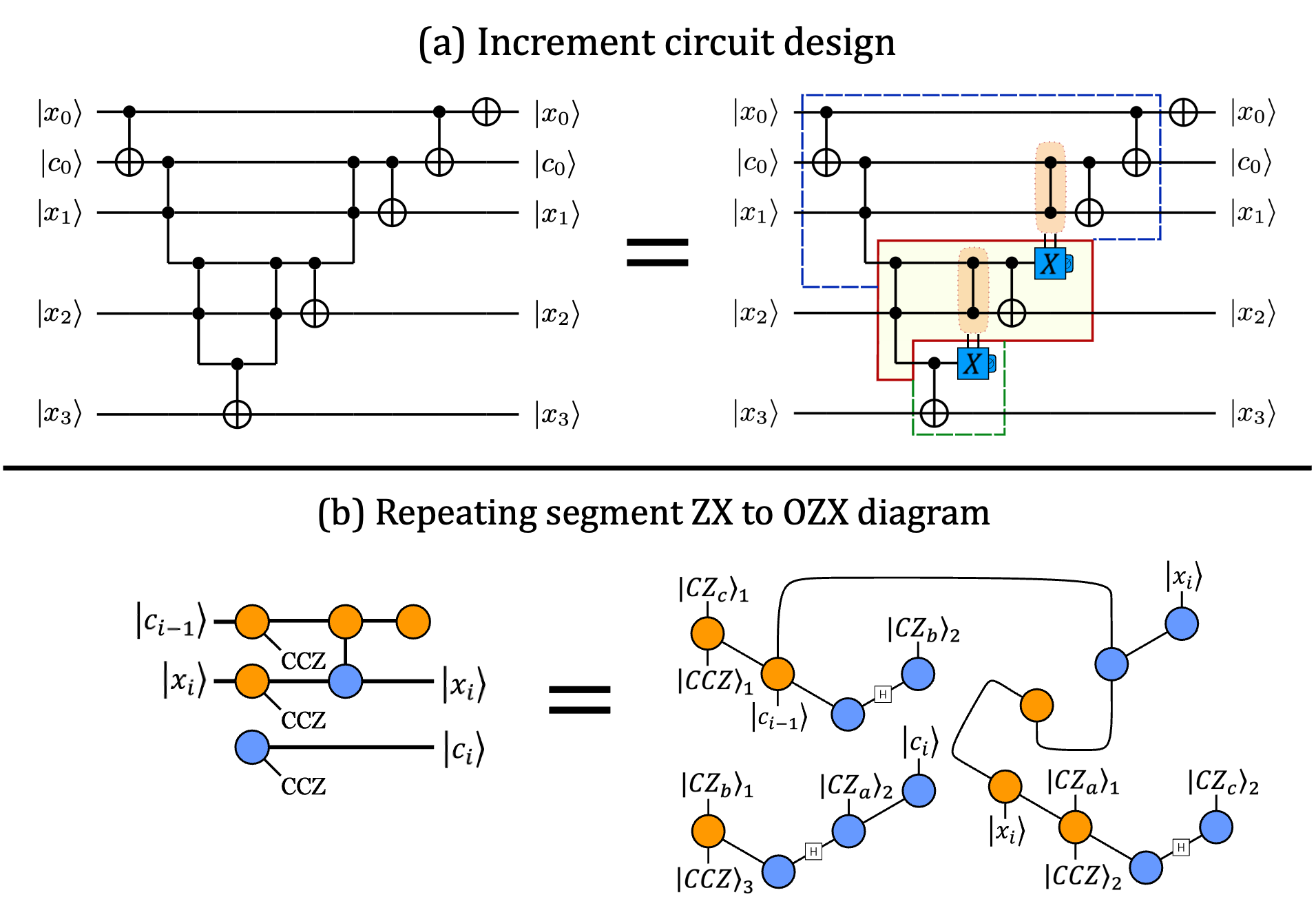}
    \caption{Here we display an example of the increment circuit on a $4$-qubit register, $\ket{x}$, where $\ket{c_{0}} = \ket{0}$ is an ancillary qubit for carry. (a) The left side presents the design in simple circuit notation; The right side shows the design with the modified temporary AND un-compute \cite{litinski2022active}. The segment with a solid red line, highlighted in yellow, represents a repeating segment. (b) The active volume calculation for a repeating segment, the block cost is $15$, the T count is $7$, and the reaction depth is $1$.}
    \label{fig:incrementor}
\end{figure*}

An efficient increment and controlled increment circuit can be produced by modifying the incrementor presented in \cite{torres_incrementor} such that each Toffoli gate pair is converted into a temporary AND. Moreover, we modify the un-compute section of the temporary AND such that it is replaced by a X basis measurement and a conditional CZ gate.
The circuit construction and the active volume calculation for the repeating segments (yellow highlighted section) is displayed in Fig.~\ref{fig:incrementor}. After including the cost of a CCZ state we find the active volume of a repeating section is $15+ C_{\ket{CCZ}}$ blocks and the baseline T count is $7$. The temporary AND un-compute commutes with all operators between it and the compute part, so the repeating section has a reaction depth of $1$. The start segment (blue dashed line) and end segment (green dashed line) have a block cost of $15 + C_{\ket{CCZ}}$ and $3$, respectively, see appendix~\ref{sec:increm_start_end}. The total active volume of an increment circuit is $(n-2)(15+C_{\ket{CCZ}})+3$, the T count is $7n-14$, and the reaction depth is $n-2$.
\par \textbf{Controlled increment.} A controlled increment can be created by changing the top two CNOTs into a temporary AND such that $\ket{x_{0}}$ and the control qubit are the controls, and the ancillary $\ket{c_{0}}$ is initialised. The cost of the first segment is now equivalent to the repeating segments such that an $n$-qubit controlled incrementor has an active volume, T count, and reaction depth of:
\begin{subequations}\label{eq:cincrem_eqns}
\begin{align}
    V_{cincrem} &=  (n-1)(15+C_{\ket{CCZ}})+3, \label{eq:cincrem_act_vol} \\
    T_{cincrem}  &= 7n-7, \label{eq:cincrem_T_count} \\
    D_{cincrem}  &= n-1. \label{eq:cincrem_depth} 
\end{align}
\end{subequations}

\subsection{Shifts \& controlled shifts}
\label{sec:shift}
In baseline architectures, a shift can be implemented by applying a chain of SWAP gates and a qubit initialisation to $\ket{0}$. The SWAP gates sequentially swap neighbouring qubits to its left or right (depending on shift direction), such that the initialised qubit is moved to the least significant bit (LSB) or most significant bit (MSB) position, respectively. This scales $O(n^2)$ where $n$ is the shift distance. 
However, for fixed-point binary arithmetic the shift can be implemented during the classical processing of the computation. This is done by simply dividing (shift right) or multiplying (shift left) the result by the $2^l$, where $l$ is the distance of the shift. 
\par In the active volume architecture shift operations are always free. Consider a fixed-point binary number $q_{n-1} q_{n-2} \ldots q_{0}$, where $q_{i} \in \{0,1\}$, each $q_{i}$ is stored in a module. These data qubits do not occupy a fixed rail as in baseline architectures, which means individual $q_{i}$ must be kept track of. Let us consider a scenario where we wish to perform a shift-by-one to the left: first, we initialise a new data qubit to state $\ket{0}$ and label it $q_0$ and then we relabel the data qubit modules such that $q_i \rightarrow q_{i+1}$. We can follow a similar process for shifts to the right and can generalise this for a $n$-distance shift. The architecture’s long-range connections therefore allow us to implement a shift with no logical block cost.
\par \textbf{Controlled shift-by-one.} Controlled shifts are conditioned on a control qubit. The state of the control qubit is determined by the algorithm and input, and therefore must be implemented quantum mechanically.
A controlled shift can be performed similarly to a regular shift, but with controlled SWAPs, where the middle CNOT of a SWAP gate is replaced by a Toffoli gate. This requires $n$ controlled SWAPs (for $n$-qubit number). 

\begin{figure}
\centering
    \includegraphics[scale=0.9]{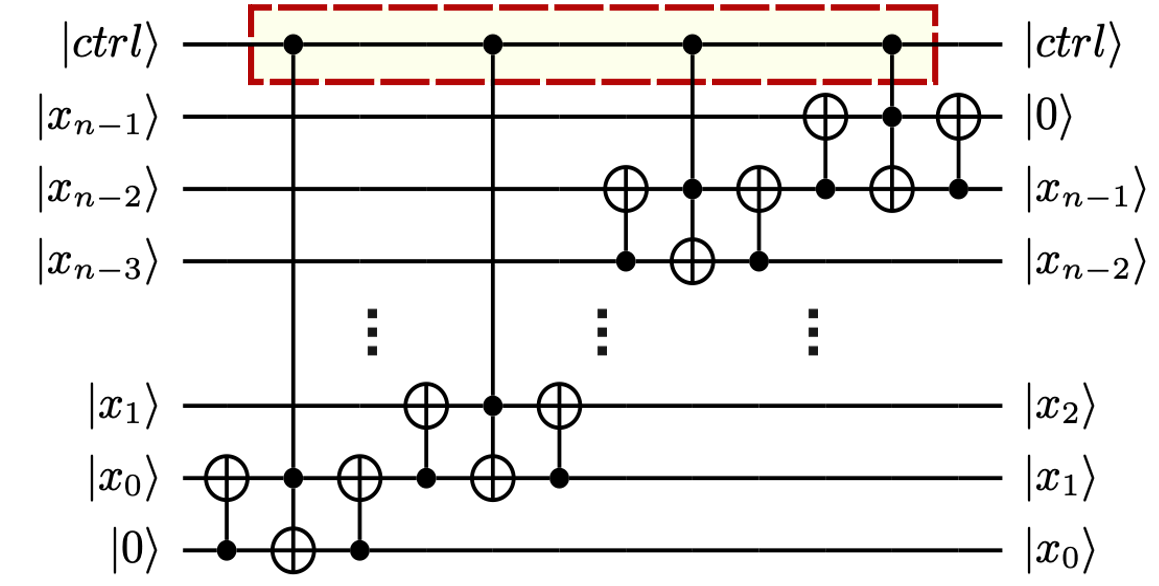}
    \caption{A controlled shift-by-one circuit. The red and dashed line box encloses the top Toffoli gate controls for each controlled SWAP. In ZX form each control is a Z spider that connects to a CCZ island, which means the block count can be reduced via spider mergers. The block cost of a controlled shift is $(20+C_{\ket{CCZ}})n$, T count $7n$, reaction depth $n$.}
       \label{fig:control_shift}
\end{figure}

Because the CNOT gates in a SWAP alternate direction (therefore spider type alternates) on each qubit rail of the target register there is no obvious ZX optimisation available. However, the control qubit exhibits a string of $n$ Z type spiders each connected to a CCZ island, Fig.~\ref{fig:control_shift}, which allows us to perform mergers. 
The number of orientated Z spiders required for $n$ Toffoli gates is given by $2 \left\lceil \frac{n-2}{4} \right\rceil +1 $, see appendix~\ref{sec:cshift_control_optimisation}. This optimisation decreases the block cost of the full operation from $(20+C_{\ket{CCZ}})n$ to $(19+C_{\ket{CCZ}})n + 2 \left\lceil \frac{n-2}{4} \right\rceil -1$, where $(20+C_{\ket{CCZ}})$ is the cost of a lone controlled SWAP. However, this reduction of $n-2 \left\lceil \frac{n-2}{4} \right\rceil +1$ is extremely small. For example, $n=32$ the block cost decreases by $15$ from approximately $1760$ to $1745$. Since the optimised expression is also more complicated, we shall use the non-optimised version. Therefore, the active volume, T count, and reaction depth of a $n$-qubit controlled shift-by-one are:
\begin{subequations}\label{eq:cshift_eqns}
\begin{align}
    V_{cshift} &=  (20+C_{\ket{CCZ}})n, \label{eq:cshift_act_vol} \\
    T_{cshift}  &= 7n, \label{eq:cshift_T_count} \\
    D_{cshift}  &= n. \label{eq:cshift_depth} 
\end{align}
\end{subequations}

\subsection{Controlled addition or subtraction}
\label{sec:CAS_adder}
Here we design a Gidney-style controlled addition or subtraction subroutine as we were unable to find any designs in literature. This operation is based around an addition circuit, and since Gidney’s adder \cite{Gidney2018adder} is the least expensive to date we expect our design to be superior to other currently possible implementations.
\par Subtraction of two binary numbers $a$ and $b$ can be implemented by applying NOT gates to register $\ket{a}$ and then performing an addition between $\ket{a}$ and $\ket{b}$, with the input carry of the LSB set to 1 \cite{litinski2023compute, cornell_twos_complement}.  
\par
\begin{figure*}[t]
\centering
    \includegraphics[scale=0.95]{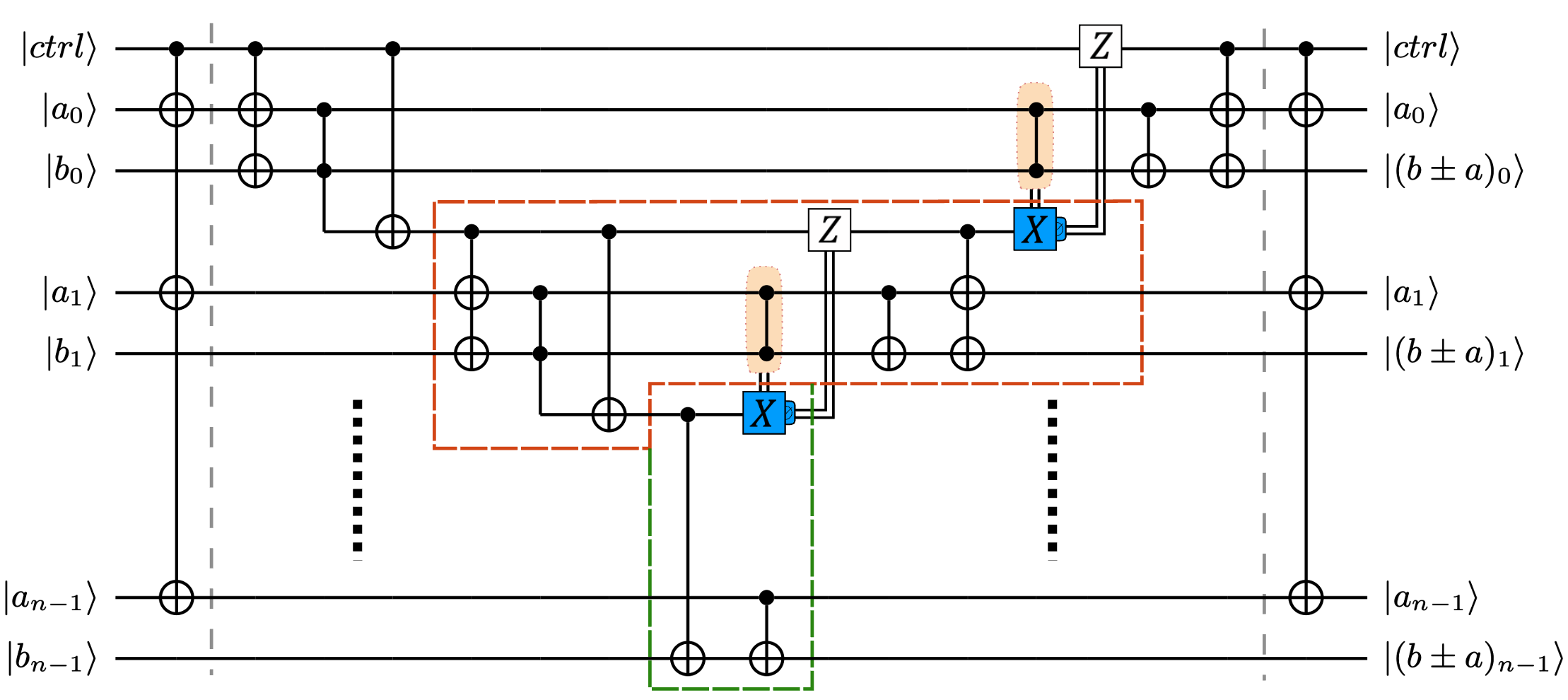}
    \caption{The circuit design of our controlled addition or subtraction (CAS) adder\footnote{The white box with a $Z$ indicates that a Pauli Z correction is required if the X measurement outcome is $-1$. This correction can be implemented by classical processing and has no effect on the cost.}. The grey dashed line divides the circuit into three sections: initial (left section) and final (right section) CNOT array, and the modified Gidney adder (middle section). The red dashed line denotes a repeating segment, while the green dashed line indicates the final segment. Both segments are identical to the repeating and final segments of the adder from \cite{litinski2022active}. The block cost of our CAS adder is $(25+C_{\ket{CCZ}})n -20 - C_{\ket{CCZ}}$, T count $7n-7$, reaction depth $2n-2$.}
       \label{fig:CAS_circuit}
\end{figure*}

 A controlled addition or subtraction transforms $\ket{a}$, $\ket{b}$, and control qubit $\ket{ctrl}$ as follows: $\ket{ctrl}\ket{a}\ket{b} \rightarrow \ket{ctrl}\ket{a}\ket{b+a}$ for $\ket{ctrl}=0$ and  $\ket{ctrl}\ket{a}\ket{b} \rightarrow \ket{ctrl}\ket{a}\ket{b-a}$ for $\ket{ctrl}=1$. This operation can be implemented using the circuit presented in Fig.~\ref{fig:CAS_circuit}, which uses $2n$ CNOT gates and a modified version of the Gidney adder from \cite{litinski2022active}. In this design, hereby referred to as a CAS adder, the control qubit is used both to perform the CNOTs on register $\ket{a}$ and as the input carry to the LSB for the addition operation. When the control qubit is 0, normal addition is performed since the input carry is 0 and $\ket{a}$ is unaffected by the CNOTs; otherwise, the subtraction method outlined in the previous paragraph is used.

\par A simplistic block calculation for the CAS circuit would be the sum of the $2n$ CNOTs, followed by the cost of the modified adder. The modified adder is the same as the adder presented in \cite{litinski2022active} but with the start segment replaced by a repeating segment. A repeating segment is defined by the red dashed red box in Fig.~\ref{fig:CAS_circuit} and the last segment is defined by the green dashed box. A repeating segment, the last segment, and a CNOT cost $(22+C_{\ket{CCZ}})$, $4$, and $4$ blocks, respectively \cite{litinski2022active}. Therefore, the na\"{i}ve approach obtains a block cost of $(30+C_{\ket{CCZ}})n -18 -C_{\ket{CCZ}}$ for the CAS Adder. However, by using the optimisation methods outlined in Section~\ref{sec:active_volume_calculation} we obtain the following costs: 
\begin{subequations}\label{eq:CAS_eqns}
\begin{align}
    V_{CAS} &=  (25+C_{\ket{CCZ}})n -20 -C_{\ket{CCZ}}, \label{eq:CAS_act_vol} \\
    T_{CAS}  &= 7n-7, \label{eq:CAS_T_count} \\
    D_{CAS}  &= 2n-2. \label{eq:CAS_depth} 
\end{align}
\end{subequations}
This is an $\approx 8\%$ reduction in active volume, for detailed calculations see appendix~\ref{sec:CAS_adder_cost}. The CAS adder requires $n-1$ ancillary qubits for carry. Note, we can construct a circuit with $n$ fewer CNOTs if we instead performed the following operation: $\ket{ctrl}\ket{a}\ket{b} \rightarrow \ket{ctrl}\ket{a}\ket{a+b}$ for $\ket{ctrl}=0$ and  $\ket{ctrl}\ket{a}\ket{b} \rightarrow \ket{ctrl}\ket{a}\ket{a-b}$ for $\ket{ctrl}=1$, see Fig.~\ref{fig:CAS_circuit_cheaper}. However, this operation is logically distinct from the '$b-a$' version above,, and so using this cheaper version requires an algorithm to be modified. 

\subsection{Multiplication}
\label{sec:multiplication}
To multiply two $n$-qubit numbers we can use repeated additions and shifts \cite{edgard2019mult, haner2018polyfit} such that, 
\begin{equation}
a \cdot b = a_{n-1}2^{n-1}b + \dots + a_{0}2^{0}b
\label{eq:addition_formula}
\end{equation}
where $a = \sum_{i} a_{i}2^{i} $ and $a_{i} \in \{ 0,1 \}$. The result register is initialised to $\ket{0}$ so like \cite{edgard2019mult} we use a Toffoli gate array to perform a controlled copy for the $i=0$ term. After this we conduct $n-1$ controlled additions, with output carry, between the $b$ register and a changing subset of qubits in the result register. 
\par
\begin{figure*}[t]
\centering
    \includegraphics[scale=0.95]{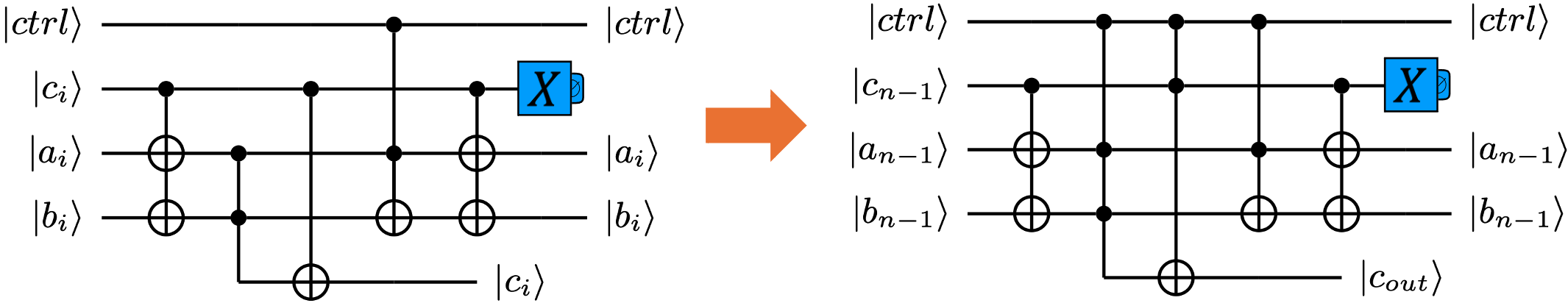}
    \caption{The left shows the design of a controlled addition repeating segment from \cite{litinski2022active}, this is also used for the repeating segments of our COG adder. The right shows a modified version of the repeating segment, which is used for the last segment of our COG adder.}
       \label{fig:COG_end_segment}
\end{figure*}

 To create a controlled adder with output carry we modify just the last segment of the controlled $n$-qubit Gidney adder presented in \cite{litinski2022active}. First, we replace the last segment with a repeating segment such that the initialised carry qubit is now the carry out. Next, we add the control qubit as a third control to the temporary AND (compute). Lastly, we promote the CNOT, acting on $\ket{c_{n-1}}$ and the carry out qubit, to a Toffoli such that the control qubit is the second control, see Fig.~\ref{fig:COG_end_segment}. This modification means when the control is 0 the carry out is $\ket{0}$ and $\ket{b_{n-1}}$ is unchanged, if the control is 1 the carry out is $\ket{ctrl(ab \bigoplus ac \bigoplus bc)_{n}}$ and $\ket{b_{n-1}} \rightarrow \ket{(b \bigoplus ctrl(a \bigoplus c))_{n-1}}$, thus a controlled addition is performed. We refer to this modified adder as a controlled overflow Gidney (COG) adder.
The block cost of a COG adder is $(30 + 2C_{\ket{\text{CCZ}}})n + 15 + 2C_{\ket{\text{CCZ}}}$ and denoted is by $V_{COGA}$, for details see appendix~\ref{sec:COG_adder_cost}. Moreover, the adder requires $n-1$ ancillary qubits and has a T count and reaction depth of $14n+14$ and $3n$, respectively. A Toffoli gate has a block cost of $12+C_{\ket{CCZ}}$ \cite{litinski2022active}. The Toffoli array has $n$ Toffoli gates and therefore costs $(12+C_{\ket{CCZ}})n$.
It follows that the block cost of a multiplication subroutine built from $n-1$ COG adders and a Toffoli array is given by:
\begin{subequations}\label{eq:mult_eqns}
\begin{align}
    V_{mult} &= (n-1) \cdot V_{COGA} + n \cdot V_{Toff} \notag \\
	      &= (30+2 C_{\ket{CCZ}})n^2 + (C_{\ket{CCZ}} \notag \\
      &\quad -3) \cdot n - 15 - 2 C_{\ket{CCZ}}, \label{eq:mult_act_vol} \\
    T_{mult}  &= (n-1) \cdot T_{COGA} + n \cdot T_{Toff} \notag \\
             	       &= 14n^2 + 7n -14, \label{eq:mult_T_count} \\
    D_{mult}  &= (n-1) \cdot D_{COGA} + n \cdot D_{Toff} \notag \\
 	        &= 3n^2 -2n, \label{eq:mult_depth} 
\end{align}
\end{subequations}
where $Toff$ labels the relevant cost of a Toffoli gate. This implementation requires $5n$ qubits, $n$ qubits for each input number, $2n$ for the result register, and $n$ ancillae.

\subsubsection{Multiplication: Baseline comparisons} 
Table~\ref{table:mult_comparison} compares the most efficient, non-approximate, design from literature, the design by Mu\~{n}oz-Coreas \& Thapliyal \cite{edgard2019mult}, against our own.
Our design beats \cite{edgard2019mult} on every metric, achieving a 62\% reduction the leading factor for T count. Note that this does not automatically mean the overall AV is smaller. However, the number of Clifford operations are similar for both constructions and so our design is obviously cheaper.

\begin{table}
\centering
\caption{Baseline cost comparison of multiplier circuits}
\small
\begin{tabular}{c||ccc}
  \hline
  \textbf{Design} & \textbf{T count} & \textbf{R Depth} & \textbf{Qubits} \\
  \hline
    \cite{edgard2019mult}   & $ 21n^2 - 14$  & $3n^2 -2$ & $6n+2$ \\
    This work   & $14n^2 + 7n -14$   & $ 3n^2 -2n $ & $5n $ \\
  \hline
\end{tabular}
\parbox[t]{\textwidth}{
\footnotesize{*R Depth is reaction depth. 
}}
\label{table:mult_comparison}
\end{table}

\subsection{Square}
\label{sec:square}
To square a number $a$ we naturally follow the same procedure used for multiplication, but we have the option to save $n-1$ qubits for the trade-off of adding $n$ CNOT gates \cite{haner2018polyfit}; 
each iteration we can copy out the qubit that is being used as the control for a controlled addition to an ancilla using a CNOT. Next, we perform the controlled addition between the $a$ register and the result register before resetting the ancilla for the next iteration using another CNOT. Note, this costs $2n$ CNOT but the standard approach still requires $n$ CNOTs to copy the input into another register, hence why we only require an additional $n$ CNOT gates. 
As we would like to reduce the qubit count, we shall consider a squaring operation which takes this trade-off. 
The cost of a squaring operation is given by:
\begin{subequations}\label{eq:square_eqns}
\begin{align}
    V_{square} &= V_{mult} + 4 \cdot 2n \notag \\
	      &= (30+2 C_{\ket{CCZ}})n^2 - 2 C_{\ket{CCZ}} \notag \\
      &\quad\quad + (5 + C_{\ket{CCZ}})n -15, \label{eq:square_act_vol} \\
    T_{square}  &= T_{mult} = 14n^2 + 7n -14, \label{eq:square_T_count} \\
    D_{square}  &= D_{mult} = 3n^2 -2n. \label{eq:square_depth} 
\end{align}
\end{subequations}

\par \textbf{Controlled square.} A controlled square subroutine is obtained by simply changing the $2n$ CNOTs into Toffoli gates:
\begin{subequations}\label{eq:ctrl_square_eqns}
\begin{align}
    V_{csquare} &= V_{mult} + 2n \cdot V_{Toff} \notag \\
	      &= (30+2 C_{\ket{CCZ}})n^2 - 2 C_{\ket{CCZ}} \notag \\
      &\quad\quad + (21 + 3C_{\ket{CCZ}})n -15, \label{eq:csquare_act_vol} \\
    T_{csquare}  &= T_{mult} + 2n \cdot T_{Toff} \notag \\
            &= 14n^2 + 21n -14, \label{eq:csquare_T_count} \\
    D_{csquare}  &= D_{mult} + 2n \cdot D_{Toff}  = 3n^2. \label{eq:csquare_depth} 
\end{align}
\end{subequations}

\section{High-level arithmetic subroutines}
\label{sec:high_level_arithmetic}

\subsection{Square root} 
\label{sec:sqrt}
The quantum square root operation has many useful applications in quantum algorithms, such as solving Pell’s equation or polynomial root finding \cite{hallgren2007, guodong2014}. It is also a component of the arcsine and logarithm subroutines explored in Sections~\ref{sec:arcsine_circuit} and \ref{sec:wang_log}. In this Section, we improve the baseline T count of the non-restoring square root algorithm circuit by Mu\~{n}oz-Coraes and Thapliyal \cite{sqrt_circuit} and determine its active volume.
\begin{figure*} [t]
\centering
\includegraphics[scale=1]{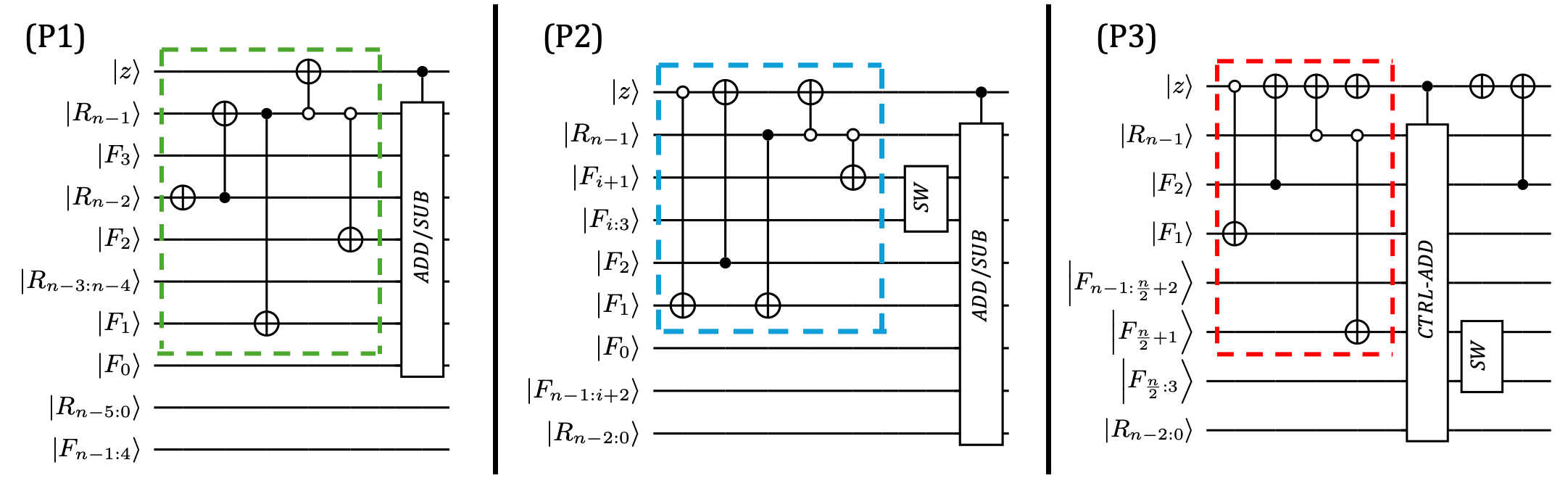} 
\caption{(P1) Circuit for the first part of the square root operation; the green box has a block cost of $10$. (P2) Circuit for the second part; the blue box has a block cost of $13$. (P3) Circuit for the third part; the red box has a block cost of $12$. Details found in appendix~\ref{sec:sqrt_cost}. “$a:b$” labels’ represent multiple qubits, $SW$ corresponds to a set of SWAP operations, $ADD/SUB$ performs an addition or subtraction conditioned on $\ket{z}$, and $CTRL-ADD$ performs an addition conditioned on $z$. This is the generalised representation of the circuit presented in \cite{sqrt_circuit}.} 
\label{fig:sqrt_P1P2P3}
\end{figure*}

\par The Mu\~{n}oz-Cores and Thapliyal square root algorithm takes an input register $\ket{a}$ and produces output registers $\ket{F}$ and $\ket{R}$. It has no garbage output and is currently the least expensive design in terms of both T count and qubits used, requiring $2n + 1$ qubits, where $n$ is the number of qubits used for the binary representation of the input $\ket{a}$ and $n$ is even. $\ket{F}$ contains the integer part of the result while $\ket{R}$ holds the remainder, both registers have $n$ qubits.
The circuit is decomposed into three parts (P1, P2, P3), see Fig.~\ref{fig:sqrt_P1P2P3}, and requires the NOT, CNOT, inverted control CNOT, SWAP, controlled addition or subtraction, and controlled addition operations. In the active volume architecture, a SWAP can be performed by simply swapping the labels of the relevant modules and applying subsequent operations appropriately. Therefore, SWAP gates have no cost and can be ignored. 
\par \textbf{P1.} Part 1 is performed once at the start of the circuit and consists of one NOT, two CNOTs, two inverted control CNOTs, and one controlled addition or subtraction on four qubits ($n=4$). The controlled ADD/SUB can be implemented using our CAS adder, detailed in Section~\ref{sec:CAS_adder}. We can optimise the first two CNOTs and inverted control CNOTs contained in the green box from (P1) Fig.~\ref{fig:sqrt_P1P2P3} such that their block cost of $16$ ($4$ blocks per CNOT \cite{litinski2022active}) is reduced to $10$, see appendix~\ref{sec:sqrt_cost}.
The logical block cost of part 1 is therefore $90+3C_{\ket{CCZ}}$, with a reaction depth of $6$, and a T count of $21$. 
\par \textbf{P2.} Part 2 is repeated $\frac{n}{2}-2$ times, and consists of two CNOTs, three inverted control CNOTs, one SWAP, and one CAS adder.
In each iteration of part 2 one of the inverted control CNOTs changes its target qubit to $\ket{F_{i+1}}$, where $i \in \{2,3,\ldots,\frac{n}{2}-1\}$. Nevertheless, we can represent the gates in the blue box from (P2) Fig.~\ref{fig:sqrt_P1P2P3} with the generalised ZX construction presented in Fig.~\ref{fig:sqrt_ZX} of the appendix as $i+1$ cannot equal $1$ or $2$ and therefore does not have spider interactions with the fixed gates acting on $\ket{F_{1}}$ and $\ket{F_{2}}$.
Optimising the blue box segment reduces it’s block cost from $20$ to $13$, see appendix~\ref{sec:sqrt_cost}.
The number of qubits that partake in the controlled addition or subtraction increases by $2$ each iteration starting with $n=6$. It follows that that sum of each part 2 operation has a block cost of $\sum^{\frac{n}{2}-1}_{i=2} V_{CAS}(2i+2) + 13$, which simplifies to $\frac{1}{4}(25+C_{\ket{CCZ}})n^2 +9n -136 -4 C_{\ket{CCZ}}$. The T count of part 2 is $\sum^{\frac{n}{2}-1}_{i=2} T_{CAS}(2i+2) = 1.75n^2 + 3.5n -14 $ and the reaction depth is $\sum^{\frac{n}{2}-1}_{i=2} D_{CAS}(2i+2) = \frac{1}{2}n^2 -8$.
\par \textbf{P3.} Part 3 is performed once at the end of the circuit and consists of two NOTs, two CNOTs, three inverted control CNOTs, two SWAP gates, and one n-bit controlled adder. Optimising the red box from (P3) Fig.~\ref{fig:sqrt_P1P2P3} reduces its block count from $16$ to $12$, see appendix~\ref{sec:sqrt_cost}.
If we use the controlled $n$-bit Gidney adder from \cite{litinski2022active} for the controlled addition, the logical block cost of part 3 is given by $(30+2C_{\ket{CCZ}})n -5 -C_{\ket{CCZ}}$, with a T count of $14n-7$, and a reaction depth of $3n-2$.
\par \textbf{Total cost.} After summing the cost of each part, we find the total logical block cost, the T count, and the reaction depth of Mu\~{n}oz-Coreas’s square root circuit to be:
\begin{subequations}\label{eq:sqrt_eqns}
\begin{align}
    V_{sqrt} &= \frac{1}{4}(25+C_{\ket{CCZ}})n^2 + (39+ \notag \\
&\quad\quad 2 C_{\ket{CCZ}})n -46 -6 C_{\ket{CCZ}}, \label{eq:sqrt_act_vol} \\
    T_{sqrt}  &= 1.75n^2 +17.5n+14, \label{eq:sqrt_T_count} \\
    D_{sqrt}  &= \frac{1}{2}n^2 +3n -4. \label{eq:sqrt_depth}
\end{align}
\end{subequations}
Note, we also require $n-1$ ancillary qubits for the CAS and controlled adders and consequently have a total qubit cost of $3n$.

\subsubsection{Square root: baseline comparison}
\begin{table*}
\centering
\caption{Baseline cost comparison of square root circuits}
\renewcommand{\arraystretch}{1.3} 
\begin{tabular}{c||cccc}
  \hline
  \textbf{Design} & \textbf{T count} & \textbf{Reaction Depth} & \textbf{Qubits} & \textbf{Circuit volume} \\
  \hline
    \cite{sqrt_circuit}   & $3.5n^2 + 21 n - 28$  & $\frac{1}{2}n^2 +3n$ & $2n+1$ & $7n^3 + 45.5n^2- 35n -28 $\\
    This work   & $1.75n^2 +17.5n+14$   & $\frac{1}{2}n^2 +3n -4 $ & $3n$ & $5.25n^3 + 52.5n^2 +42n$\\
  \hline
\end{tabular}
\label{table:sqrt_comparison}
\end{table*}
Table~\ref{table:sqrt_comparison} compares our implementation against the original, \cite{sqrt_circuit}, which is the leading design in literature \cite{sultana_sqrt, sqrt_circuit, ananthalakshmi2017novel}. We ignore the T depth reported in \cite{sqrt_circuit} and calculate their reaction depth in the same manner as we do for our own circuits, which gives the upper bound on reaction depth. 
The controlled adder used by \cite{sqrt_circuit}, \cite{edgard2019mult}, has a Toffoli count of $3n+2$, a T count of $21n+14$, and reaction depth $3n+2$. Their controlled addition or subtraction \cite{thapliyal2016addsub}, has a Toffoli count of $2n-2$, a T count of $14n-14$, and reaction depth $2n-2$. Both adders are either more expensive or equal to our own in every category and we see the total baseline cost reflect this.
One drawback of our design is that it requires $n-1$ more qubits but, as the overall circuit volume is still reduced and active volume computing nullifies the cost of idle qubits, we consider this a worthwhile trade-off.
To conclude our design is overall less expensive than \cite{sqrt_circuit} and so we use its active volume to represent the block cost of a square root operation.

\subsection{Trigonometry: Piecewise polynomial evaluation (PPE)}
\label{sec:poly_eval_circuit}
H\"{a}ner et al. created a circuit with a T count $O(n^2)$ for evaluating polynomials that can be used to approximate convergent transcendental functions \cite{haner2018polyfit}. Here we modify and then calculate the active volume H\"{a}ner’s circuit.
This method can be used to evaluate $\sin{x}$, $\cos{x}$, $\arcsin{x}$, $\arccos{x}$, $\exp{-x}$, $\exp{-x^2}$, $\tanh{x}$, or any other function that is well approximated by polynomials.
Cao et al. \cite{cao2013} also proposed a method based on Taylor expansion to evaluate $\sin{x}$ and $\arcsin{x}$. Like H\"{a}ner's approach, their $\sin{x}$ implementation has a computational complexity of $O(n^2)$ in quantum operations. However, Cao's approach for $\arcsin{x}$ is more costly as it requires multiple invocations of $\sin{x}$. Moreover, the method has not been implemented at the Toffoli gate level and does not offer the flexibility to evaluate a range of functions, which H\"{a}ner's method does. 
A third approach by Wang et al. \cite{wang2020tran} can accurately approximate trig functions and provides an exact evaluation of their inverses but is more expensive requiring $O(n^3)$ quantum operations ($O(n^3)$ T gates).
We will however revisit Wang’s work in Section~\ref{sec:wang_log} to implement a $\log{x}$ subroutine. $\log{x}$ is divergent function around $x=0$ and therefore polynomial approximation fails.
\par H\"{a}ner’s piecewise polynomial approximation circuit is based on the classical Horner scheme which evaluates a polynomial $P(x)$ expressed as,
\begin{equation} 
P(x) = \sum^{q}_{i=0} a_{i} x^{i},
\label{eq:horner_scheme} 
\end{equation}
where $q$ is the degree of the polynomial and $a_{i}$ is the $i^{th}$ coefficient of the polynomial $i \in \{q,q-1,\ldots,0\}$.
H\"{a}ner uses a label register to allow for case distinctions based on which subdomain the input value of $x$ falls into. Then using the label register the circuit evaluates a $k$-degree polynomial that best approximates $f(x)$ in that subdomain. 
The full piecewise polynomial evaluation (PPE) circuit consists of a label finding circuit and polynomial evaluation circuit. The label finding circuit is composed of $M$ comparisons and CNOT gates. 
The polynomial evaluation circuit comprises $q$ iterations each containing one multiplication, one addition, and a $NEXT$ operation, which loads the next set of classically determined polynomial coefficients.
While the multiplication subroutine as described in Section~\ref{sec:multiplication} can be used in the iterative step, we follow H\"{a}ner’s approach and use a cheaper method for fixed-point multiplication that also keeps the result from growing beyond $n$ qubits.  
The method introduces a total error per multiplication of $\varepsilon = \frac{n}{2^{n-p}}$, where the input $x$ is given in binary fixed-point representation and $p$ is the number of qubits to the left of the fixed point. This error is a factor of $n$ greater than the error produced using the standard multiplier \cite{haner2018polyfit}.
The cheap multiplication involves $n$ additions on $1,2,3,\ldots,n$ qubits. The additions are performed similarly to full multiplication, but only on the $n$ most significant qubits of the result register.
Note, it could be useful to consider bridge qubits for the controlled additions such that the effective depth of the multiplication is reduced by parallelisation. 

\begin{figure}
\hspace{-1.1cm}
\includegraphics[scale=0.8]{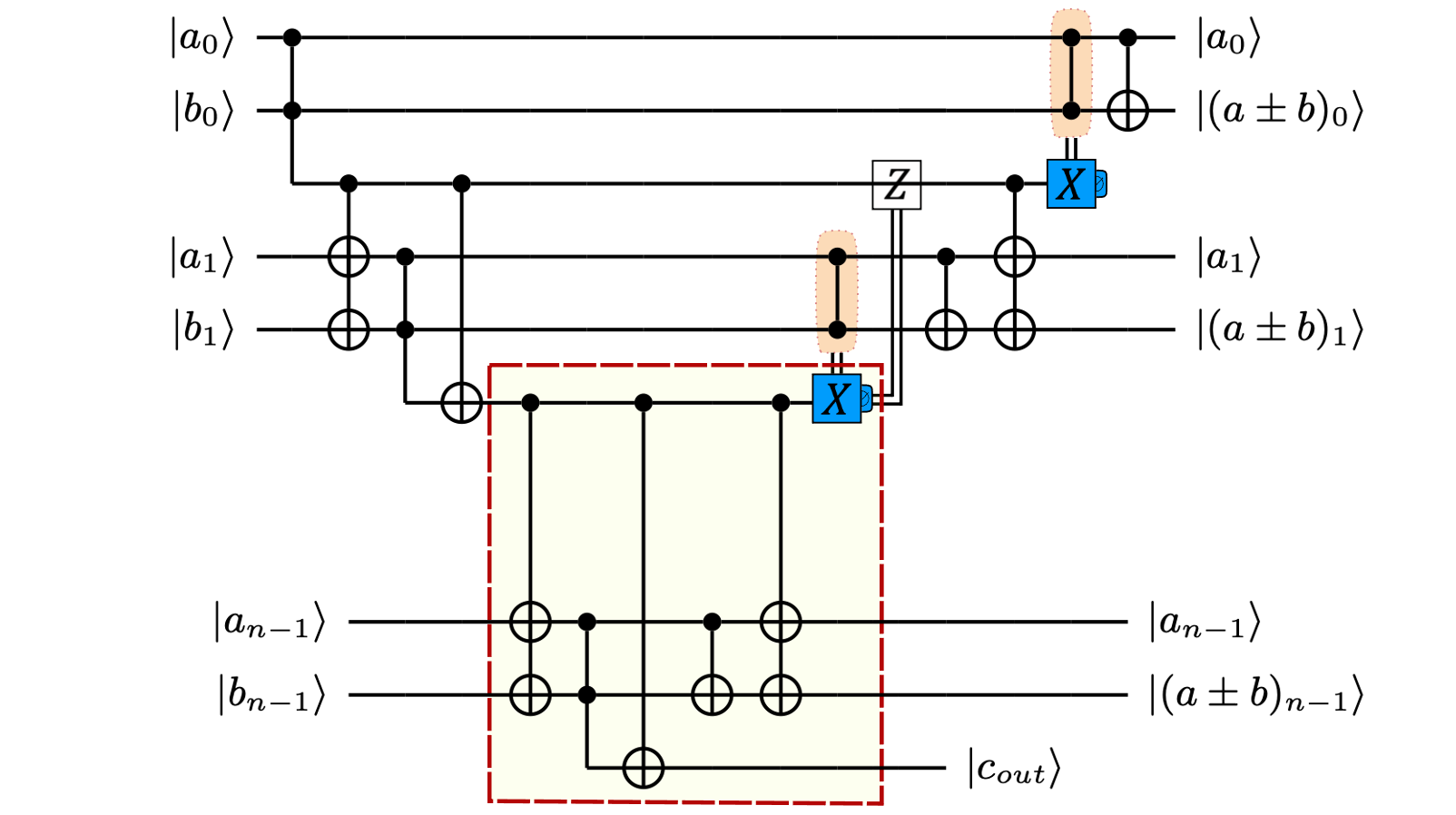} 
\caption{The circuit for our Gidney-styler adder with output carry (OG adder). The circuit is identical to the adder from \cite{litinski2022active} apart from the last segment which is highlighted in the red box. The cost of the final segment is the same as a repeating segment, i.e. $22+C_{\ket{CCZ}}$ \cite{litinski2022active}. The active volume of the OG adder is $(22+C_{\ket{CCZ}})n -7$, the T count is $7n$, the reaction depth is $2n-2$, see appdendix~\ref{sec:OG_adder_cost}} 
\label{fig:OG_adder_circuit}
\end{figure}

\par
\begin{figure}
\hspace{-0.4cm}
\includegraphics[scale=0.45]{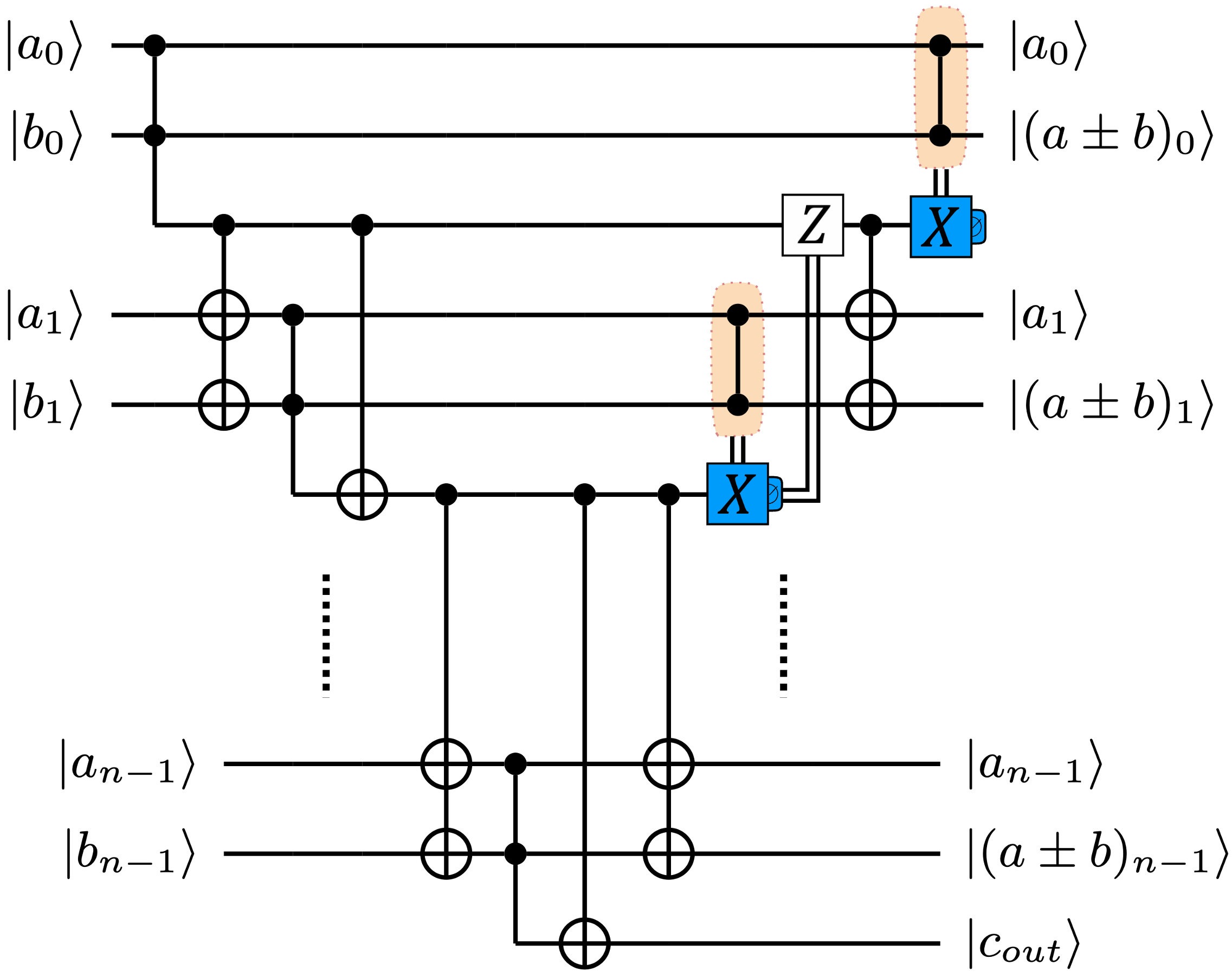} 
\caption{The circuit for our Gidney-styler comparator. The circuit is similar to the adder from \cite{litinski2022active}, but with the CNOTs between $a$ and $b$ removed and the final segment replaced with a repeating segment. The active volume of the comparator is $(20+C_{\ket{CCZ}})n -8$, the T count is $7n$, the reaction depth is $2n-1$, see appendix~\ref{sec:cmp_cost}} 
\label{fig:cmp_circuit}
\end{figure}

 \textbf{Baseline cost reduction.} Before calculating the total costs, we can reduce the baseline T count by introducing a few modifications. 
Firstly, we look to replace the adders used by H\"{a}ner, where an addition costs $2n-1$ Toffoli gates (T count $14n-7$, reaction depth $2n-1$) and a controlled addition costs $3n+3$ Toffoli gates (T count $21n+21$, reaction depth $3n+3$) \cite{takahashi2009adder}.
The addition operation used in the polynomial approximation circuit allows for output carry, which is not handled in the standard Gidney adder circuit. 
To remedy this, we created the modified Gidney adder presented in Fig.~\ref{fig:OG_adder_circuit}, hereby referred to as an OG adder (output-carry Gidney adder).
The OG adder is a modified version of the adder presented in \cite{litinski2022active}, where the last segment has a cost that is equivalent to a repeating segment. 
The total block cost of an OG adder is $(22+C_{\ket{CCZ}})n-7$, the T count is $7n$, and the reaction depth is $2n-2$, for details see appendix~\ref{sec:OG_adder_cost}.
Our next modification alters the multiplication operator. We replace the multiplier used by H\"{a}ner, which is based on the controlled adder from \cite{takahashi2009adder}, with our Toffoli array/COG adder multiplier presented in Section~\ref{sec:multiplication}.
Thirdly, we replace the H\"{a}ner’s square root method with our subroutine adaptation of \cite{sqrt_circuit} as this method far less expensive and reduces the total error. This gives a T count reduction of $187.25n^2+269.5n+336np+336p-336p^2-56$ and a reaction depth reduction of $\frac{53}{2}n^2+38n+48np+48p-48p^2-2$, see appendix~\ref{sec:haner_sqrt_cost_comparison} for details. One drawback of this substitution is that we require an extra $n$ qubit register to complete the calculation but the T count reduction more than compensates for this.
Finally, we construct our own OG adder-based comparator in place of the original, which uses the CARRY circuit from \cite{haner2017comparator}, giving a T count of $56n-84$ and reaction depth $8n-12$. To construct our own comparator, we first note: a comparison between two numbers $x$ and $y$ can be achieved by finding the one’s complement of $x$ (apply X gates to each qubit in $x$) and then performing an addition with $y$. If the carry output is $1$ then $y>x$ \cite{how_to_make_comparator}. 
However, this would require us to uncompute the addition of $x$ and $y$ after copying out the carry result. Instead, we modify our OG adder so only one ladder of temporary-AND gates are needed, see Fig.~\ref{fig:cmp_circuit}.
The cost of a comparison is derived in appendix~\ref{sec:cmp_cost}: 
\begin{subequations}\label{eq:cmp_eqns}
\begin{align}
    V_{cmp} &= (20+C_{\ket{CCZ}})n-8 \label{eq:cmp_act_vol} \\
    T_{cmp} &= 7n, \label{eq:cmp_T_count} \\
    D_{cmp} &= 2n-1, \label{eq:cmp_depth}
\end{align}
\end{subequations}
Therefore, this modification reduces the T count and reaction depth of a comparison by $49n-84$ and $6n-4$, respectively.
\par \textbf{PPE circuit costs.} First will shall consider the active volume of the cheap multiplication method.
Each iteration, the output carry of the addition operation becomes the LSB of the multiplication operation’s result register in the next iteration. 
A consequence of this is that we can no longer use a Toffoli array for the first step of a multiplication. Instead, this array must be substituted for a controlled addition. 
This means the total block cost, T count, and reaction depth of the cheap fixed-point multiplication is given by:
\begin{subequations}\label{eq:cmult_eqns}
\begin{align}
    V_{mult}^{(c)} &= \sum^{p-1}_{i=0} V_{COGA}(n-i) + \sum^{n-p}_{i=1} V_{COGA}(n-i) \notag \\
	&= (30+2C_{\ket{CCZ}}) (\frac{n^2}{2}-\frac{n}{2}+np+p-p^2) \notag \\
&\quad\quad + (15+ 2C_{\ket{CCZ}})n, \label{eq:cmult_act_vol} \\
    T_{mult}^{(c)}  &= \sum^{p-1}_{i=0} T_{COGA}(n-i) + \sum^{n-p}_{i=1} T_{COGA}(n-i) \notag \\
 	&= 7n^2 +7n+14np+14p-14p^2, \label{eq:cmult_T_count} \\
    D_{mult}^{(c)}  &= \sum^{p-1}_{i=0} D_{COGA}(n-i) + \sum^{n-p}_{i=1} D_{COGA}(n-i) \notag \\
&= \frac{3}{2}n^2 -\frac{3}{2}n +3np + 3p -3p^2, \label{eq:cmult_depth}
\end{align}
\end{subequations}
where we have obtained the initial expressions with the summations by trivially adapting the Toffoli count equation in \cite{haner2018polyfit}. Next, we calculate the cost of a fused multiplication and addition by adding the cost of an $n$-qubit OG adder, $V_{OGA}$:
\begin{subequations}\label{eq:fma_eqns}
\begin{align}
    V_{fma} &= V_{mult}^{(c)} + V_{OGA} \notag \\
	&= (30+2C_{\ket{CCZ}}) (\frac{n^2}{2}-\frac{n}{2}+np+p-p^2) \notag \\
&\quad\quad + (37+ 3C_{\ket{CCZ}})n -7, \label{eq:fma_act_vol} \\
    T_{fma}  &= T_{mult}^{(c)} + T_{OGA} \notag \\
 	&= 7n^2 +14n+14np+14p-14p^2, \label{eq:fma_T_count} \\
    D_{fma}  &= D_{mult}^{(c)} + D_{OGA} \notag \\
&= \frac{3}{2}n^2 +\frac{1}{2}n +3np + 3p -3p^2 -2. \label{eq:fma_depth}
\end{align}
\end{subequations}
To evaluate a $q$-degree polynomial we need $q$ multiplications and additions, Eq.~\ref{eq:horner_scheme}, therefore the cost of the polynomial evaluation circuit is: 
\begin{subequations}\label{eq:poly_eqns}
\begin{align}
    V_{poly} &= q \cdot V_{fma} \notag \\
	&= q(30+2C_{\ket{CCZ}}) (\frac{n^2}{2}-\frac{n}{2}+np+p-p^2) \notag \\
&\quad\quad + (37+ 3C_{\ket{CCZ}})qn -7q, \label{eq:poly_act_vol} \\
    T_{poly}  &= q \cdot T_{fma} \notag \\
 	&= 7qn^2 +14qn+14qnp \notag \\
&\quad \quad +14qp-14qp^2, \label{eq:poly_T_count} \\
    D_{poly}  &= q \cdot D_{fma} \notag \\
&= \frac{3}{2}qn^2 +\frac{1}{2}qn +3qnp \notag \\
&\quad \quad + 3qp -3qp^2 -2q. \label{eq:poly_depth}
\end{align}
\end{subequations}

\begin{figure}
\hspace{1.cm}
\includegraphics[scale=0.8]{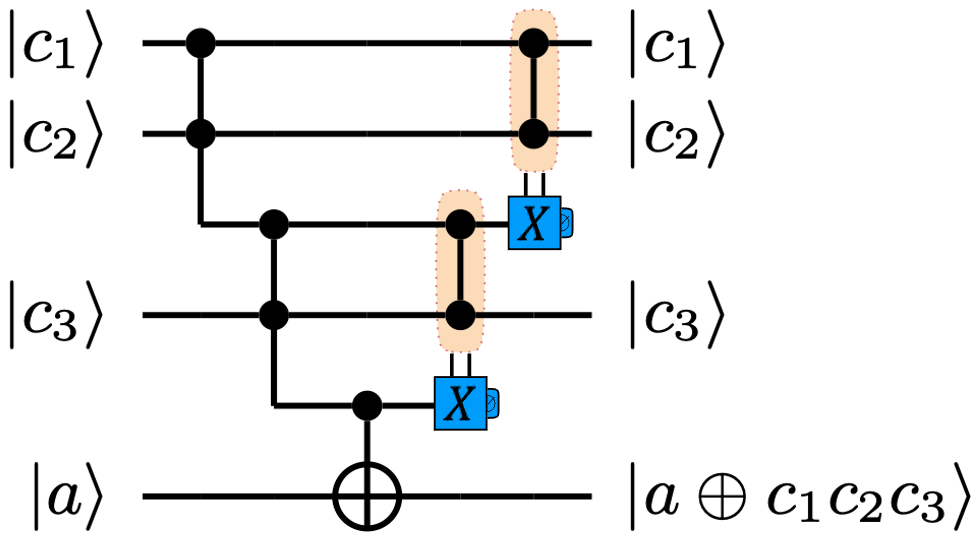} 
\caption{A circuit which implements a $k$-controlled NOT gate using $k-1$ temporary AND gates. It has an active volume of $(k-1)(12+C_{\ket{CCZ}})+3$, T count $7k-7$, reaction depth $k-1$, and requires $k-1$ ancillary qubits.}
\label{fig:k_controlled_NOT}
\end{figure}

The circuit evaluates $M$ polynomials at a time using a superposition of all label states and a $NEXT$ operation, which loads the $i^{th}+1$ set of polynomial coefficients. This operation is implemented with $2M$ $\left\lceil \log_2 M \right\rceil$-controlled NOT gates \cite{haner2018polyfit}. 
We can construct a $k$-controlled NOT gate using $k-1$ temporary AND gates and $k-1$ ancillary qubits, as illustrated in Fig.~\ref{fig:k_controlled_NOT}. Each temporary AND costs the same as a Toffoli gate and there are $k-1$ of them. The CNOT in the last segment is the same as an incrementor end segment, Fig.~\ref{fig:increm_start_end_cost}, and costs $3$ blocks. Therefore, a $k$-controlled NOT has active volume $V_{kNOT}=(k-1)(12+C_{\ket{CCZ}})+3$, with a T count of $7k-7$, and reaction depth $k-1$.
In contrast, \cite{haner2018polyfit} performs a $k$-controlled NOT using $4(k-2)$ Toffoli gates, which has a T count of $28(k-2)$ and a reaction depth of $4(k-2)$. Therefore, our construction reduces the T count and reaction depth of a $k$-controlled NOT by approximately three-quarters. The cost of our $NEXT$ operation is given by:
\begin{subequations}\label{eq:next_eqns}
\begin{align}
    V_{next} &= 2M \cdot V_{kNOT}(\left\lceil \log_2 M \right\rceil) \notag \\
	&= 2M(\left\lceil \log_2 M \right\rceil -1)(12+ \notag \\
&\quad \quad C_{\ket{CZZ}})+6M, \label{eq:next_act_vol} \\
    T_{next}  &= 2M \cdot T_{kNOT} \notag \\
&= 14M(\left\lceil \log_2 M \right\rceil -1), \label{eq:next_T_count} \\
    D_{next}  &= 2M \cdot D_{kNOT} \notag \\
&= 2M(\left\lceil \log_2 M \right\rceil -1). \label{eq:next_depth}
\end{align}
\end{subequations}
The label register is composed of $M$ comparisons and controlled increments.
The controlled increments are conditioned on the output carry of the comparator such that each comparison will increment the label register until input $x \leq \eta$, where $\eta$ is the lower boundary of the next subdomain. 
H\"{a}ner points out that the controlled increments can be performed cheaply with CNOTs on the label register sparing us the cost of a generalised controlled increment operation \cite{haner2018polyfit}. However, the paper does not provide details and so we derive our own expression:
The number of CNOTs depends on number of qubits in the label register, $\left\lceil\log_2 M \right\rceil$. Let $\ket{l}=\ket{0}$ be the first subdomain of the label register. Then it can be shown, see appendix~\ref{sec:label_CNOTs}, that the number of CNOTs required to increment from $\ket{l}=\ket{0}$ to  $\ket{l}=\ket{M}$ is upper bounded by $2^{z+1} -z -2$, where $z=\left\lceil\log_2 M \right\rceil $. Therefore, the cost of the label finding circuit is:
\begin{subequations}\label{eq:label_eqns}
\begin{align}
    V_{label} &= M \cdot V_{cmp}+ 4(2^{\left\lceil \log_2 M \right\rceil +1} -z -2) \notag \\
	&= M(20+C_{\ket{CZZ}})n -8M - 8  \notag \\
&\quad \quad + 2^{\left\lceil \log_2 M \right\rceil +3}-4\left\lceil \log_2 M \right\rceil, \label{eq:label_act_vol} \\
    T_{label}  &= M \cdot T_{cmp} =  7Mn, \label{eq:label_T_count} \\
    D_{label}  &= M \cdot D_{cmp} = 2Mn-M, \label{eq:label_depth} 
\end{align}
\end{subequations}
where $cmp$ refers to the relevant cost of our comparator. The total cost of the piecewise polynomial evaluation (PPE) circuit is given by:
\begin{subequations}\label{eq:PPE_eqns}
\begin{align}
    V_{PPE} &= V_{poly} + q \cdot V_{next} + V_{label} = q(15 \notag \\
&\quad +C_{\ket{CZZ}}) (\frac{n^2}{2}-\frac{n}{2}+np+p-p^2) \notag \\
&\quad +(37+3C_{\ket{CZZ}})qn -7q+6Mq  \notag \\
&\quad + Mq(\left\lceil \log_2 M \right\rceil -1)(24 +2C_{\ket{CZZ}}) \notag \\
&\quad +M(20+C_{\ket{CZZ}})n -8M \notag \\
	&\quad + 2^{\left\lceil \log_2 M \right\rceil +3} +4\left\lceil \log_2 M \right\rceil -8, \label{eq:PPE_act_vol} \\
    T_{PPE}  &= T_{poly} + q \cdot T_{next} + T_{label} = 7(qn^2 \notag \\
&\quad +3.5qnp+3.5qp-3.5qp^2) +14qn \notag \\
&\quad +[14Mq(\left\lceil \log_2 M \right\rceil -1) +7Mn], \label{eq:PPE_T_count} \\
    D_{PPE}  &= D_{poly} + q \cdot D_{next} + D_{label} = \frac{3}{2}qn^2 \notag \\
&\quad +\frac{1}{2}qn +3qnp +3qp -3qp^2 -2q \notag \\
&\quad \quad\quad+2Mq(\left\lceil \log_2 M \right\rceil-1) \notag \\
&\quad \quad\quad + 2Mn-M. \label{eq:PPE_depth} 
\end{align}
\end{subequations}
For a reversible implementation we must store the output of each fused multiplication and addition iteration. In total we require $(q+2)n+ \left\lceil \log_2 M \right\rceil + s$ qubits for the PPE circuit. That is for the $q$ intermediates value registers, one result register, one label register, one comparator result qubit, $s$ for the polynomial coefficient register, and $n-1$ ancillary qubits for the adders and multipliers.

\subsubsection{PPE: baseline \& active volume comparison} 
\begin{table*}
\centering
\caption{Baseline cost comparison of PPE circuits}
\label{table:PPE_comparison}
\renewcommand{\arraystretch}{1.3} 
\begin{tabular}{c||ccc}
  \hline
  \textbf{Design} & \textbf{T count} & \textbf{Reaction Depth} & \textbf{Qubits} \\
  \hline
    \cite{haner2018polyfit}   & $84n^2 +658n +10710$ & $\frac{27}{2}n^2 +94n +1530$ & $7n+5$ \\
    This work   & $63n^2 +318.5n +4032$  & $\frac{27}{2}n^2 +76n +500$ & $8n+4$ \\
    \hline
\end{tabular}
\parbox[t]{\textwidth}{
\hspace{2.9cm}
\footnotesize{*$M=16$, $p=\frac{n}{2}$, $q=6$ in this comparison.}
}
\end{table*}
The original PPE circuit has the following T count and reaction depth: 
\begin{subequations}\label{eq:PPE_eqns2}
\begin{align}
    T_{PPE}^{\text{H\"{a}ner}}  &= 10.5(qn^2+3.5qnp+3.5qp-3.5qp^2) \notag \\
&\quad +[56Mq(\left\lceil \log_2 M \right\rceil -2) +28Mn] \notag \\
&\quad \quad \quad+24.5nq -7q, \label{eq:PPE_T_count_H} \\
    D_{PPE}^{\text{H\"{a}ner}}  &= \frac{3}{2}qn^2 +\frac{7}{2}qn +3qnp+3qp-3qp^2 \notag \\
&\quad +2Mq(4\left\lceil \log_2 M \right\rceil -8) \notag \\
&\quad \quad \quad+4Mn -q. \label{eq:PPE_D_count_H} 
\end{align}
\end{subequations}
These expressions were obtained by reasoning that the reaction depth is equal to Toffoli count equation in \cite{haner2018polyfit} and the T count is this equation multiplied by $7$.
The T count of our modified implementation, as shown in Eq.~\ref{eq:PPE_T_count}, significantly reduces the T count compared to Eq.~\ref{eq:PPE_T_count_H}. Specifically, there is a $2/3$ reduction in the terms within the first set of brackets. Moreover, the first term and the second term within the square brackets are reduced by factors of $4$ (approximately) and $2$, respectively. The remaining terms also decrease in cost, transforming from $24.5nq-7q$ to $14nq$. In table~\ref{table:PPE_comparison} we provide a clear comparison with realistic parameter settings by setting $M=16$, $p=\frac{n}{2}$, and $q=6$.

\subsection{Inverse trigonometry: arcsine}
\label{sec:arcsine_circuit}
We can implement a quantum arcsine via a combination of polynomial evaluation and a square root, this follows from the classical method used in the maths library Cephes \cite{haner2018polyfit, moshier2000cephes}. 
The method relies on case distinction: case 1, where $|x| < 0.5$, and case 2, where $0.5 \le x \le 1$. For case 1 arcsine can be well approximated by a single polynomial; For case 2 we use the identity given in Eq.~\eqref{eq:case2eqn}, which requires us to compute a square root. 
This identity allows us to evaluate $\arcsin{x}$ away from $x = \pm 1$, where $\arcsin{x}$ diverges and cannot be well approximated by polynomials.
\begin{equation} 
\arcsin{(x)}=\frac{\pi}{2}-2\arcsin{(\sqrt{z})}, \label{eq:case2eqn} \end{equation}
where $z=\frac{1-x}{2}$ and satisfies $z \in [0,0.25]$ for $x \in [0.5,1]$. 
\cite{haner2018polyfit} describes a circuit to implement this case based arcsine method, but here we present a simpler adaptation: The circuit takes input register $\ket{x}$, where $\ket{x_{n-1}}$ is the MSB and the pseudo sign bit, and returns $\ket{x} \ket{\arcsin{x}}$. We split the circuit into six parts:
\newline \\
\textbf{Part 1:}
The PPE circuit requires the input to be positive \cite{haner2018polyfit} and so in part 1 we perform a conditional 2’s complement. 
\begin{itemize}
\item Step 1: Initialise an ancillary qubit to state $\ket{0}$, let this qubit be labelled $\ket{\alpha}$.
\item Step 2: Apply a CNOT gate such that $\ket{x_{n-1}}$ is the control and $\ket{\alpha}$ is the target.
\item Step 3: For $j=0$ to $j=n-1$, where $j$ increases in integer steps of $1$, apply a CNOT gate such that $\ket{\alpha}$ is the control and $\ket{x_{j}}$  is the target.
\item Step 4: Apply a controlled increment such that $\ket{\alpha}$ is the control and $\ket{x}$ is conditionally incremented by $1$. 
\end{itemize}
\textit{Output:} 
\begin{equation} \ket{\alpha} \ket{|x|} \notag
\end{equation}
To simplify expressions, we shall now assume that $x$ is positive such that $|x|=x$.

\textbf{Part 2:}
Here we create a register that holds $\sqrt{z}$ and a initialise a qubit whose value reveals whether we are in case 1 or case 2. As $x \in [0,1]$ we can use an inverted control temporary AND to determine if $x>0.5$ \cite{haner2018polyfit}. 
\begin{itemize}
\item Step 1: Apply an inverted control temporary AND (compute) such that the controls $\ket{x_{n-2}}$ and $\ket{x_{n-3}}$ initialise an ancillary qubit. This ancillary qubit will be labelled as $\ket{\beta}$. $\ket{\beta}=\ket{0}$ if $\ket{x_{n-2}}=\ket{x_{n-3}}=\ket{0}$ else $\ket{\beta}=\ket{1}$.
\item Step 2: Initialise an $n$-qubit register to state $\ket{1}$, let this this register be labelled $\ket{\sqrt{z}}$.
\item Step 3: Apply a subtraction between $\ket{x}$ and $\ket{\sqrt{z}}$ such that $\ket{\sqrt{z}}$ holds the result and becomes $\ket{\sqrt{z}}=\ket{1-x}$.
\item Step 4: Apply a shift-R on $\ket{\sqrt{z}}$ such that $\ket{\sqrt{z}}=\ket{\frac{1-x}{2}}$.
\item Step 5: Apply a square root on $\ket{\sqrt{z}}$ such that $\ket{\sqrt{z}}=\ket{\sqrt{\frac{1-x}{2}}}$.
\end{itemize}
\textit{Output:}
\begin{equation} \ket{\alpha} \ket{\beta} \ket{x} \ket{ \sqrt{z}} \notag 
\end{equation}

\textbf{Part 3:}
Here we conditionally load either $x$ or $\sqrt{\frac{1-x}{2}}$ into the polynomial evaluation circuit to obtain either $\arcsin{x}$ or $\arcsin{\sqrt{\frac{1-x}{2}}}$.
\begin{itemize}
\item Step 1: For $j=0$ to $j=n-1$, where $j$ increases in integer steps of $1$, apply a mixed control temporary AND (compute) such that control $\ket{\sqrt{z}_{j}}$ and inverted control $\ket{\beta}$ initialise $\ket{j}$ of an $n$-qubit ancillary register. Let the ancillary register be labelled $\ket{P_{in}}$. This step conditionally initialises register $\ket{P_{in}}$ to $\ket{\sqrt{z}}$ if $\ket{\beta}=\ket{0}$ or to $\ket{0}$ if $\ket{\beta}=\ket{1}$.
\item Step 2: Apply a controlled copy such that $\ket{\beta}$ is the control and $\ket{x}$ is conditionally copied to $\ket{P_{in}}$. After this step $\ket{P_{in}}=\ket{\sqrt{z}}$ if $\ket{\beta}=\ket{0}$ or $\ket{P_{in}}=\ket{x}$ if $\ket{\beta}=\ket{1}$.
\item Step 3: Initialise $q$ $n$-qubit registers.
\item Step 4: Apply a polynomial evaluation circuit, where the polynomial coefficients construct an approximation for $\arcsin{x}$ on the interval $[0,0.5]$, such that $\ket{P_{in}}$ is the input and one of the $q$ $n$-qubit registers from step 3 holds the result. Let the result register be labelled $\ket{P_{out}}$. The other $q-1$ $n$-qubit registers from step 3 will hold the intermediate results.
\end{itemize}
After steps 1-4 $\ket{P_{out}} =\ket{\arcsin{x}}$ if $\ket{\beta}=\ket{1}$ or $\ket{P_{out}} =\ket{\arcsin{\sqrt{z}}}$ if $\ket{\beta}=\ket{0}$
\newline \textit{Output:} 
\begin{equation} \ket{\alpha} \ket{\beta} \ket{x} \ket{\sqrt{z}} \ket{P_{in} } \ket{P_{out}} \ket{\chi} \notag
\end{equation}
$\ket{\chi}$ denotes the registers of the intermediate results.

\textbf{Part 4:}
Here we create the final output conditioned on whether we are in case 1 or 2. 
\begin{itemize}
\item Step 1: Apply an inverted control shift-L such that $\ket{\beta}$ is the inverted control and $\ket{P_{out}}$ is conditionally shifted to the left. 
\item Step 2: Initialise an $n$-qubit register. Let this register be labelled $\ket{R}$
\item Step 3: Apply a series of inverted CNOT gates such that $\ket{\beta}$ is the inverted control and $\ket{R}$ is the target. If $\ket{\beta}=\ket{0}$ then $\ket{R}=\ket{\frac{\pi}{2}}$ otherwise there is no change. 
\item Step 4: Apply an inverted control addition or subtraction such that $\ket{\beta}$ is the inverted control and register $\ket{P_{out}}$ and $\ket{R}$ are conditionally added or subtracted with $\ket{R}$ holding the result. 
\end{itemize}
After steps 1-4 $\ket{R}=\ket{\frac{\pi}{2}-2\arcsin{\sqrt{z}}}$ if $\ket{\beta}=\ket{0}$ or $\ket{R}=\ket{\arcsin{x}}$ if $\ket{\beta}=\ket{1}$.
\newline \textit{Output:} 
\begin{equation} 
\ket{\alpha} \ket{\beta} \ket{x} \ket{\sqrt{z}} \ket{P_{in} } \ket{P_{out}} \ket{R} \ket{\chi} \notag
\end{equation}

\textbf{Part 5:}
Here we un-compute all unneeded qubits.
\begin{itemize}
\item Step 1: Apply an inverted control shift-R such that $\ket{\beta}$ is the inverted control and $\ket{P_{out}}$ is conditionally shifted to the right. 
\item Step 2: Run the polynomial evaluation circuit in reverse such that $\ket{P_{out}}$ and $\ket{\chi}$ are un-computed back to $\ket{0}$.
\item Step 3: Apply a controlled copy such that $\ket{\beta}$ is the control and register $\ket{x}$ is conditionally un-copied from register $\ket{P_{in}}$.
\item Step 4: For $j=0$ to $j=n-1$, where $j$ increases in integer steps of $1$, apply a mixed control temporary AND (un-compute) such that control $\ket{\sqrt{z}_{j}}$ and inverted control $\ket{\beta}$ remove qubit $\ket{P_{in}^{j}}$. This step removes register $\ket{P_{in}}$.
\item Step 5: Perform part 2 in reverse such that register $\ket{\sqrt{z}}$ is returned to $\ket{0}$ and qubit $\ket{\beta}$ is removed. 
\end{itemize}
\textit{Output:} 
\begin{equation} 
\ket{\alpha} \ket{x} \ket{R} \notag
\end{equation}

\textbf{Part 6:}
Here we correct the sign bit such that it matches the original input $x$ and remove the last ancilla $\ket{\alpha}$.
\begin{itemize}
\item Step 1: For $j=0$ to $j=n-1$, where $j$ increases in integer steps of $1$, apply two CNOT gates, where $\ket{\alpha}$ is the control for both, $\ket{x_{j}}$ is the target of one, and $\ket{R_{j}}$ is the target of the other. 
\item Step 2: Apply a controlled increment such that $\ket{\alpha}$ is the control and $\ket{x}$ is conditionally incremented by $1$. 
\item Step 3: Apply a controlled increment such that $\ket{\alpha}$ is the control and $\ket{R}$ is conditionally incremented by $1$.
\item Step 4: Apply a CNOT such that $\ket{x}$ is the control and $\ket{\alpha}$ is the target. This returns qubit $\ket{\alpha}$ to $\ket{0}$. 
\end{itemize}
\textit{Output:} 
\begin{equation} 
 \ket{x} \ket{\arcsin{x}} \notag 
\end{equation}

\par \textbf{Arcsine circuit costs.} Part 1 requires $n+1$ CNOT gates and a controlled increment. 
Part 2 requires one temporary AND (compute) gate and one square root. 
Part 3 requires $n$ temporary AND (compute) gates, $n$ Toffoli gates, and one polynomial evaluation circuit. 
Part 4 requires one controlled shift, the controlled initialisation of $\frac{\pi}{2}$, and one controlled addition or subtraction. As $\frac{\pi}{2}$ is irrational it is impossible to know how many CNOTs are required for this step until a value for $n$ is chosen, so we shall overestimate the cost and use $n$ CNOT gates for this step. 
Part 5 requires a controlled shift, one polynomial evaluation circuit, $n$ Toffoli gates, $n+1$ temporary AND (un-compute) gates, and one square root.
Part 6 requires $2n+1$ CNOTs and two controlled increments. 
The sum of each part gives the cost of the arcsine circuit:
\begin{subequations}\label{eq:arcsin_eqns}
\begin{align}
V_{arcsin} &= 2V_{poly} + 2V_{sqrt} + 2V_{CAS} +2n \cdot V_{Toff} \notag \\
&\quad + (2n+1) \cdot (V_{tmpAND} + V_{tmpAND}^{'}) \notag \\
&\quad + 2V_{cshift} + 3V_{cincrem}  \notag \\ 
&= q(30+2C_{\ket{CCZ}} )(n^2 +n + 2np + 2p \notag \\ 
&\quad - 2p^2) + q(74+6C_{\ket{CCZ}})n + \frac{1}{2}(25 \notag \\
&\quad +C_{\ket{CCZ}})n^2 + (256+12C_{\ket{CCZ}})n \notag \\
&\quad  -14q -126 -15C_{\ket{CCZ}}, \label{eq:arcsine_act_vol}
\end{align}
where $V_{tmpAND}=9+C_{\ket{CCZ}}$ is the active volume of a temporary AND compute, and $V_{tmpAND}^{'}=5$ is the active volume of a temporary AND un-compute \cite{litinski2022active}. The T count and reaction depth are given by:
\begin{align}
T_{arcsin} &= 2T_{poly} + 2T_{sqrt} + 2T_{CAS} +2n \cdot T_{Toff} \notag \\
&\quad + (2n+1) T_{tmpAND} + 2T_{cshift} + 3T_{cincrem}  \notag \\ 
&= 14qn^2 +28qn + 28qnp + 28qp \notag \\
&\quad -28qp^2 + 3.5n^2 +105n +7, \label{eq:arcsine_T_count} \\
D_{arcsin} &= 2D_{poly} + 2D_{sqrt} + 2D_{CAS} +2n \cdot D_{Toff} \notag \\
&\quad + (2n+1) (D_{tmpAND} + D_{tmpAND}^{'}) \notag \\
&\quad + 2D_{cshift} + 3D_{cincrem}  \notag \\ 
&= 3qn^2 +qn + 6qnp + 6qp - 6qp^2 \notag \\
&\quad - 4q + n^2 + 19n -11. \label{eq:arcsine_depth}
\end{align}
\end{subequations}
To perform an $\arcsin$ operation we require $(q+5)n + 2$ qubits (this includes ancillaries addition etc). 

\subsubsection{Arcsine: baseline \& active volume comparison}
\begin{table*}
\centering
\caption{Baseline cost comparison of arcsine circuits}
\renewcommand{\arraystretch}{1.3} 
\begin{tabular}{c||ccc}
  \hline
  \textbf{Design} & \textbf{T count} & \textbf{Reaction Depth} & \textbf{Qubits} \\
  \hline
    \cite{haner2018polyfit}   & $750.75n^2 + 1477 n - 154$  & $\frac{429}{2}n^2 +211n -22$  & $9n+7$ \\
    This work   & $129.5n^2 +294n +7$   & $28n^2 +43n -35 $ & $11n+2$ \\
  \hline
\end{tabular}
\parbox[c]{\textwidth}{
\hspace{3.0cm}
\footnotesize{*$m=3$, $p=\frac{n}{2}$, and $q=6$ in this comparison.} \\
}
\label{table:arcsine_T_count_comparison}
\end{table*}
The Toffoli count for H\"{a}ner’s arcsine implementation is provided as an equation in \cite{haner2018polyfit}. Since Toffoli gates have a reaction depth of $1$, the circuit depth is directly given by this equation. The T count of the circuit is obtained by multiplying the given expression by $7$.
\begin{subequations} \label{eq:arcsin_H_eqns}
\begin{align}
T_{arcsin}^{\text{(H\"{a}ner)}} &= q(21n^2 + 49n + 42np +42p  \notag \\
&\quad  - 42p^2 - 14) + 388.5n^2 +710.5n \notag \\ 
&\quad + 693np +693p - 693p^2 -70, \label{eq:haner_arcsine_T_count} \\
D_{arcsin}^{\text{(H\"{a}ner)}} &= q(3n^2 + 7n + 6np+6p - 6p^2 - 2) \notag \\
&\quad  + \frac{111}{2}n^2 +\frac{203}{2}n + 99np \notag \\
&\quad \quad +99p - 99p^2 - 10. \label{eq:haner_arcsine_depth}  
\end{align}
\end{subequations} 
To obtain these expressions the number of Newton-Raphson iterations was set to $m=3$. Note that the parameter $m$ only appears in H\"{a}ner’s implementation and is due to their square root method. The T count reduction observed when comparing our expression Eq.~\eqref{eq:arcsine_T_count} to H\"{a}ner’s Eq.~\eqref{eq:haner_arcsine_T_count} is not immediately clear. Therefore, we provide table~\ref{table:arcsine_T_count_comparison}, where $p=\frac{n}{2}$ and $q=6$. Our baseline implementation has an almost $6$-fold reduction in the $n^2$ term for the T count, but increases the qubit count by $2n-5$.  

\subsection{Logarithm}
\label{sec:wang_log}
The PPE circuit cannot be used to directly implement divergent functions (at least not across their entire domain) as polynomial approximation fails around divergent points. One option to evaluate logarithmic functions is the method presented by Bhaskar et al. \cite{bhaskar2015}. The method involves Newton iteration and repeated square roots but requires user-set parameters to manage error propagation, this means its implementation cost is difficult generalise. Instead, we use a method by Wang et al. \cite{wang2020tran}. Wang introduces the quantum function-value binary expansion (qFBE) method that is similar to the classical CORDIC system \cite{cordic1959}. The qFBE method outputs $\log{x}$ digit-by-digit through a recursive process and can also be used to evaluate exact inverse trigonometric functions. However, it is $O(n^3)$ in quantum operations and $O(n^2)$ in qubits, which is a power of $n$ greater than the piecewise polynomial approximation method from Section~\ref{sec:poly_eval_circuit}.
\par The qFBE method takes in an $m$-qubit input $a_{0}$ in domain $I=[1,2)$ and has $n-1$ iterations. Each iteration uses an intermediary state $a_{i}$ whose MSB is copied to the output register. Since inputs can exist in a superposition of states that may not all align with the domain $I$, a shift (left or right) is performed to ensure that the minimum value in the superposition falls within the domain. The number of positions shifted is denoted by $v$. Additionally, we require a parameter $h$ such that the difference between the minimum and maximum values in the superposition are $2^h-1 \le \Delta y \le 2^{h+1} -1$. Wang only presents a circuit implementation for the simplest case where $v=0$ and $h=1$. Here we adapt Wang’s circuit and obtain the generalised form of $\log{x}$. We can also make this circuit reversible by the storing intermediate states $\ket{a_{1:n-1}}$.
\par \textbf{Log circuit costs.} Conditioned on the MSB of $a_{i}$ we perform either $(a_{i})^2$ or $(a_{i})^2 /2^{h+1}$ each iteration. For our reversible implementation we compute the square of $a_{i}$ into a $2m$ register, then copy out the first $m$ bits into a new register which will store $a_{i+1}$. Then we un-compute the square resetting the result register for the next iteration.
We can reduce the number of controlled shift-R operations by performing $k$ controlled shift-R operations before the square operation, where $k=\left\lfloor \frac{h+1}{2} \right\rfloor$. This employs the following equivalence:
\begin{equation}
(a_{i})^2/2^{h+1} = (a_{i}/2^k)^2/2^{(h+1) \mod 2}.
\end{equation}
Per iteration we perform $k+(h+1) \mod 2$ controlled shift-Rs, two $m$-qubit square operations, and $m+1$ CNOTs.
After completing the recursive process, we multiply by $h+1$. The number of controlled additions in the multiplication depends on $\alpha$, the number of qubits in the binary representation of $h+1$. We require $\alpha-1$ controlled additions and one Toffoli array such that the cost of this multiplication is:
\begin{subequations} \label{eq:mult_log_eqns}
\begin{align}
V_{mult}^{\log} &= (\alpha -1)V_{COGA}+ n \cdot V_{Toff} \notag \\
&= [(27+2C_{\ket{CCZ}}) \alpha -23 \notag \\ 
&\quad -2C_{\ket{CCZ}}]n + (18+2C_{\ket{CCZ}})\alpha \notag \\
&\quad \quad -18 -2C_{\ket{CCZ}}, \label{eq:mult_log_act_vol} \\
T_{mult}^{\log} &= (\alpha -1)T_{COGA}+ n \cdot T_{Toff} \notag \\
&= (14\alpha -7)n + 14(\alpha-1), \label{eq:mult_log_T_count} \\
D_{mult}^{\log} &= (\alpha -1)T_{COGA}+ n \cdot T_{Toff} \notag \\
&= 3\alpha n -2n. \label{eq:mult_log_depth}  
\end{align}
\end{subequations} 
Note, if $h=1$ then this final multiplication can be replaced by a shift, i.e multiply by $2$. Next, we must add or subtract $v$ depending on whether the initial shift was to the left or right. This can be done with our CAS adder by utilising an ancillary qubit as the control whose value reflects the direction of the shift. If $v=0$ then this addition can be ignored. Finally, we shall set $m=n$, as simulations have shown that level of truncation for the intermediate register provides an exact evaluation \cite{wang2020tran}. 
In total a $\log$ operation requires $(n-1)(k+(h+1) \mod{2})$ $n$-qubit controlled shifts, $2(n-1)$ square operations, $n^2-1$ CNOTs, one (log) multiplication, and one n-bit CAS adder. Therefore, the active volume, T count, and reaction depth of the $\log$ subroutine are:
\begin{subequations} \label{eq:log_eqns}
\begin{align}
V_{log} &= (n -1)(\beta \cdot V_{cshift}+ 2 V_{square} +4n+4) \notag \\
	&\quad +V_{mult}^{\log}+ V_{CAS} = (60+4C_{\ket{CCZ}})n^3 \notag \\
	&\quad -(46+6C_{\ket{CCZ}} -(20+C_{\ket{CCZ}})\beta)n^2 \notag \\
	&\quad -(41+7C_{\ket{CCZ}}+(20+C_{\ket{CCZ}})\beta \notag \\
&\quad - (30 +2C_{\ket{CCZ}})\alpha )n + (15+ \notag\\
&\quad\quad 2C_{\ket{CCZ}})\alpha -9-3C_{\ket{CCZ}},  \label{eq:log_act_vol} \\
T_{log} &= (n -1)(\beta \cdot T_{cshift}+ 2 T_{square}) \notag \\
	&\quad +T_{mult}^{\log}+ T_{CAS} = 28n^3 \notag \\
&\quad +(14+7\beta)n^2 -(98+7\beta -14\alpha)n \notag \\
&\quad \quad +14\alpha +35, \label{eq:log_T_count} \\
D_{log} &= (n -1)(\beta \cdot D_{cshift}+ 2 D_{square}) \notag \\
	&\quad +D_{mult}^{\log}+ D_{CAS} = 6n^3 -(10 +\beta)n^2  \notag \\
	&\quad \quad +(4-\beta +3\alpha)n-2, \label{eq:log_depth}  
\end{align}
\end{subequations} 
where $\beta = k + (h+1) \mod 2$. The circuit requires $n$ qubits for the input, $2(n-1)n$ qubits for the intermediate states, $n$ qubits for the output, an $n$ qubit register to store $v$ (in binary), and $n$ qubits for the square operation’s ancillaries (we reuse these qubits for the multiplication and the addition/subtraction in the last step). The total number of qubits required is $2n^2+2n$. Wang does not provide a baseline cost for their implementation and so we are unable to provide a baseline cost comparison.

\subsubsection{Method limitations \& improvements}
A key limitation of Wang’s method is that there is currently no circuit to determine $l$ or $v$. Therefore, these parameters must be classically determined, which may not always be feasible. To reduce the cost of this log implementation, one could try replacing the square subroutine with the square root subroutine run in reverse. The square root subroutine has a significantly smaller T count and active volume compared to the square subroutine. However, this might lead to large error accumulation due to truncation, so an analysis on error propagation would be necessary. Utilising the cheap multiply method for the square operations is another option that would reduce the qubit requirements and active volume, each intermediary state would now require $n$ instead of $2n$ qubits. However, again further analysis of error propagation is needed. Lastly, if we expect the inputs to be away from the divergence at $\log{(x=0)}$, then it is cheaper to use the piecewise polynomial approximation circuit.

\section{Conclusion}
We have significantly reduced the T counts of various arithmetic functions compared to their original implementations, mainly by making use of the temporary AND gate, see table~\ref{fig:baseline_all_subroutines}. 
Secondly, we have expanded the number of optimised subroutines for active volume computing to include several high-level arithmetic functions such as $\sin{x}$, $\arcsin{x}$, and $\log{x}$, see table~\ref{fig:all_subroutines}. Furthermore, we provide a simple methodology for deriving optimised active volumes. We see that circuit designs which maximise uninterrupted lines of controls and targets naturally lead to reduced block counts due to simple spider mergers. 
\par An interesting and perhaps unexpected finding (shown only in Fig.~\ref{fig:COOP_adder_no_overflow_end_compute}) is that logically equivalent circuits with the same gate counts can have different active volumes due to the circuit structure alone. This sensitivity to structural arrangement opens a new optimisation consideration for active volume computers that is not captured when only considering gate counts. This finding not only underscores the utility of the active volume model but also suggests that AV-aware circuit design may reveal hidden inefficiencies in otherwise “optimised” circuits.
\par \textbf{Future work and outlook.} Other useful low-level arithmetic subroutines that have not yet been analysed include carry look-ahead adders and a division circuit \cite{draper2004adder, thapliyal2017divide}. The optimisation and inclusion of these circuits would complete the set of low-level arithmetic subroutines. One could also attempt to further reduce the cost of operations by investigating conjugacy classes through ZX and circuit level manipulations, see Fig.~\ref{sec:example_act_vol_cost}. However, with the expanded toolkit provided by this work, active volume resource estimates for important algorithms that use trigonometric functions, such as the HHL algorithm, can now be conducted. The results of such research may prove to be more insightful about the benefits of the architecture. In conclusion, we hope this paper will prompt further investigation into active volume computing and encourage future work to not only report Clifford gate counts but also to consider Clifford gate optimisation.

\section*{Acknowledgments}
I would like to thank Terry Rudolph for introducing me to the concept of active volume, guiding the direction of this research, feedback on drafts, and most importantly for helpful discussions on relevant topics and ideas. I would also like to thank Florian Mintert for volunteering to supervise the masters thesis on which this paper is an adaption. 

\bibliographystyle{ieeetr}
\bibliography{references.bib}

\onecolumn \newpage
\appendix

\section{Summary tables \& simulation code}
\label{sec:table_summaries}

\subsection{Circuit simulation code}
The simulation code for all the low-level circuits can be found here: 
\href{https://github.com/HEAVEYS/Quantum-Circuits.git}{GitHub Repository: Quantum-Circuits}.

\begin{figure}[h]
\small
\renewcommand{\arraystretch}{1}
 \begin{adjustwidth}{}{}
\begin{tabular}{cccc}
\textbf{Subroutine} & \textbf{T count} & \textbf{Peak qubits} & \textbf{Section} \\
\hline
\multicolumn{4}{c}{Low-level arithmetic} \\ 
\hline
\multicolumn{1}{l}{k-controlled NOT} & $7k-7$ & $2k$ & \ref{sec:poly_eval_circuit} \\
\multicolumn{1}{l}{Increment} & $7n-14$ & $2n-1$ & \ref{sec:incrementors} \\
\multicolumn{1}{l}{Controlled Increment} & $7n-7$ & $2n$ & \ref{sec:incrementors} \\
\multicolumn{1}{l}{Controlled shift-by-one} & $7n$ & $n+2$ & \ref{sec:shift} \\
\multicolumn{1}{l}{Addition (with output carry)} & $7n$ & $3n$ & \ref{sec:poly_eval_circuit} \\
\multicolumn{1}{l}{Controlled addition (with output carry)} & $14n+17$ & $3n+1$ & \ref{sec:multiplication} \\
\multicolumn{1}{l}{Controlled addition or subtraction} & $7n-7$ & $3n$ & \ref{sec:CAS_adder} \\
\multicolumn{1}{l}{Multiplication} & $14n^2 +7n-14$ & $5n$ & \ref{sec:multiplication} \\
\multicolumn{1}{l}{Square} & $14n^2+7n-14$ & $4n$ & \ref{sec:square} \\
\multicolumn{1}{l}{Controlled square} & $14n^2+21n-14$
 & $4n+1$ & \ref{sec:square} \\
\multicolumn{1}{l}{Comparison} & $7n$ & $3n$ & \ref{sec:poly_eval_circuit} \\
\hline
\multicolumn{4}{c}{High-level arithmetic} \\ 
\hline
\multicolumn{1}{l}{Square Root} & $1.75n^2+17.5n+14$
 & $3n+1$  & \ref{sec:sqrt} \\
\multicolumn{1}{l}{Polynomial evaluation} & $7qn^2 +14qn+14qnp+14qp-14qp^2$ & $(q+2)n+s$ & \ref{sec:poly_eval_circuit} \\
\multicolumn{1}{l}{Piecewise polynomial evaluation} & $7qn^2 +14qn+14qnp+14qp-14qp^2$ & $(q+2)n+\left\lceil \log_2 M \right\rceil $ & \ref{sec:poly_eval_circuit} \\ & $+ 14Mq(\left\lceil \log_2 M \right\rceil -1)+7Mn$ & $+ s$ & \\
\multicolumn{1}{l}{$\arcsin(x)$} & $14qn^2 + 28qn + 28qnp + 28qp$ & $(q+5)n+2$ & \ref{sec:arcsine_circuit} \\ & $-28qp^2 +3.5n^2 +105n +7$ & & \\
\multicolumn{1}{l}{$\log{(x)}$} & $28n^3 + (14+7\beta)n^2 -(98+7\beta$ & $2n^2+2n$ &\ref{sec:wang_log} \\ & $- 14\alpha)n + 14\alpha +35$ & & \\ 
\end{tabular}
\caption{\label{fig:baseline_all_subroutines} The T count and number of ancillary qubits for all of the subroutines mentioned in this paper. $q$ denotes the degree of the polynomial, $p$ is the number of fractional digits in the binary fixed-point representation of the input registers, $M$ is the number of segments in the piecewise polynomial evaluation circuit, $s$, is the number of qubits used for polynomial coefficients $a_i$, and the definitions of $\alpha$ and $\beta$ are found in Section~\ref{sec:wang_log}. The reaction depths are displayed in Fig~\ref{fig:all_subroutines}. Note by 'Peak qubits' we mean the maximum number of qubits required during an operation in order for it to be implemented.}
\end{adjustwidth}
\end{figure}

\begin{figure}[p]
\small
\renewcommand{\arraystretch}{1.5}
 \begin{adjustwidth}{-1.65cm}{}
\begin{tabular}{cccc}
\textbf{Subroutine} & \textbf{Active Volume (in blocks)} & \textbf{Reaction Depth} & \textbf{Section} \\
\hline
\multicolumn{4}{c}{Low-level arithmetic} \\ 
\hline
\multicolumn{1}{l}{k-controlled NOT} & $(12+C_{\ket{CCZ}})(k-1)+3$ & $k-1$ & \ref{sec:poly_eval_circuit} \\
\multicolumn{1}{l}{Increment} & $(15+C_{\ket{CCZ}})(n-2)+3$ & $n-2$ & \ref{sec:incrementors} \\
\multicolumn{1}{l}{Controlled Increment} & $(15+C_{\ket{CCZ}})(n-1)$ & $n-1$ & \ref{sec:incrementors} \\
\multicolumn{1}{l}{Controlled shift-by-one} & $(20+C_{\ket{CCZ}})n$ & $n$ & \ref{sec:shift} \\
\multicolumn{1}{l}{Addition (with output carry)} & $(22+C_{\ket{CCZ}})n -7$ & $2n-1$ & \ref{sec:poly_eval_circuit} \\
\multicolumn{1}{l}{Controlled addition (with output carry)} & $(30+2C_{\ket{CCZ}})(n+1)-15$ & $3n$ & \ref{sec:multiplication} \\
\multicolumn{1}{l}{Controlled addition or subtraction} & $(25+C_{\ket{CCZ}})(n-1) +5$ & $2n-2$ & \ref{sec:CAS_adder} \\
\multicolumn{1}{l}{Multiplication} & $(30+2 C_{\ket{CCZ}})n^2 + (C_{\ket{CCZ}}-3)\cdot n - 15 - 2 C_{\ket{CCZ}}$ & $3n^2 -2n$ & \ref{sec:multiplication} \\
\multicolumn{1}{l}{Square} & $(30+2 C_{\ket{CCZ}})n^2 + (5+C_{\ket{CCZ}})n - 15 - 2 C_{\ket{CCZ}}$ & $3n^2 -2$ & \ref{sec:square} \\
\multicolumn{1}{l}{Controlled square} & $(30+2 C_{\ket{CCZ}})n^2 + (21 + 3C_{\ket{CCZ}})n -15 - 2 C_{\ket{CCZ}}$
 & $3n^2$ & \ref{sec:square} \\
\multicolumn{1}{l}{Comparison} & $(20+C_{\ket{CCZ}})n-8$ & $2n-1$ & \ref{sec:poly_eval_circuit} \\
\hline
\multicolumn{4}{c}{High-level arithmetic} \\ 
\hline
\multicolumn{1}{l}{Square Root} & $\frac{1}{4}(25+C_{\ket{CCZ}})n^2 + (39+2 C_{\ket{CCZ}})n -46 -6 C_{\ket{CCZ}}$
 & $\frac{1}{2}n^2 +3n -4$  & \ref{sec:sqrt} \\
\multicolumn{1}{l}{Polynomial evaluation} & $q(30+2C_{\ket{CCZ}}) (\frac{n^2}{2} -\frac{n}{2}+np+p-p^2)$ & $q(\frac{3}{2}n^2 +\frac{1}{2}n +3np$ & \ref{sec:poly_eval_circuit} \\ & $+ (37+ 3C_{\ket{CCZ}})qn -7q$ & $+ 3p -3p^2 -2)$ & \\
\multicolumn{1}{l}{Piecewise polynomial evaluation} & $q(15+C_{\ket{CZZ}}) (\frac{n^2}{2}-\frac{n}{2}+np+p-p^2) +(37$ & $q(\frac{3}{2}n^2  +\frac{1}{2}n +3np$ & \ref{sec:poly_eval_circuit} \\  & $+3C_{\ket{CZZ}})qn-7q+6Mq + Mq(\left\lceil \log_2 M \right\rceil -1)\cdot$ & $+ 3p -3p^2 -2)$ & \\
& $(24 +2C_{\ket{CZZ}})+M(20+C_{\ket{CZZ}})n -8M+$ & $+2Mq(\left\lceil \log_2 M \right\rceil$ & \\ & $2^{\left\lceil \log_2 M \right\rceil +3} +4\left\lceil \log_2 M \right\rceil -8$ & $-1)+2Mn-2$ & \\
\multicolumn{1}{l}{$\arcsin(x)$} & $q(30+2C_{\ket{CCZ}} )(n^2 +n + 2np + 2p- 2p^2)$ & $q(3n^2 +n + 6np $ & \ref{sec:arcsine_circuit} \\ & $+ q(74+6C_{\ket{CCZ}})n + \frac{1}{2}(25+C_{\ket{CCZ}})n^2$ & $+ 6p- 6p^2- 4)$ & \\
& $+ (256+12C_{\ket{CCZ}})n -14q -126 -15C_{\ket{CCZ}} $ & $+ n^2 + 19n -11$ & \\
\multicolumn{1}{l}{$\log{(x)}$} & $(60+4C_{\ket{CCZ}})n^3-(46+6C_{\ket{CCZ}} -(20$ & $6n^3 -(10 +\beta)n^2$ &\ref{sec:wang_log} \\ & $+C_{\ket{CCZ}})\beta)n^2-(41+7C_{\ket{CCZ}}+(20+C_{\ket{CCZ}})\beta -$ & $+(4-\beta +3\alpha)n$ & \\ & $(30 +2C_{\ket{CCZ}})\alpha )n + (15+2C_{\ket{CCZ}})\alpha -9-3C_{\ket{CCZ}} $ & $-2$& \\
\end{tabular}
\caption{\label{fig:all_subroutines} The active volumes and reaction depths of all subroutines mentioned in this paper. $q$ denotes the degree of the polynomial, $p$ is the number of fractional digits in the binary fixed-point representation of the input registers, $M$ is the number of segments in the piecewise polynomial evaluation circuit.}
\end{adjustwidth}
\end{figure}

\FloatBarrier

\section{Active volume calculations}
\label{sec:av_details}

\subsection{Example of circuit compilation effecting active volume}
\label{sec:example_act_vol_cost}
\begin{figure*}[h]
\centering
\includegraphics[scale=0.95]{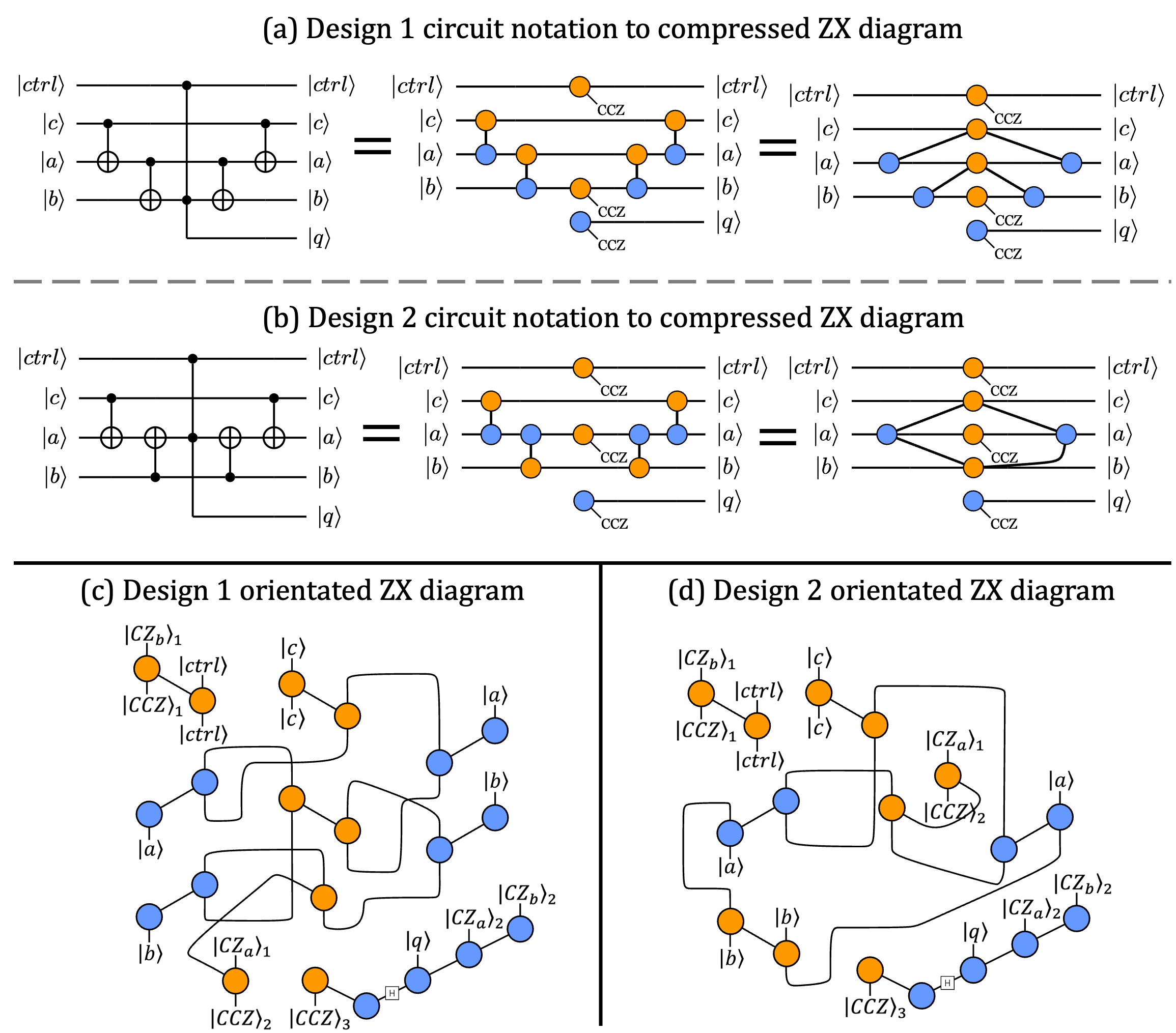} 
\caption{Two designs for the controlled out-of-place addition of three qubits (compute) with no overflow. $\ket{ctrl}$ is the control qubit and $\ket{c}$, $\ket{a}$, and $\ket{b}$ are the inputs, $q=ctrl \cdot (a \bigoplus b \bigoplus c)$ is the addition’s result. Both circuits are composed of the same gates, however, design 1 costs $21$ blocks while design 2 costs $17$. Note the qubit labels $a$, $b$, and $c$ are not related to the $a$, $b$, and $c$ labels on the $CZ$ states}
\label{fig:COOP_adder_no_overflow_end_compute}
\end{figure*}

\subsection{Increment start and end segments}
\label{sec:increm_start_end}

\begin{figure}[h]
\centering
\includegraphics[scale=1.0]{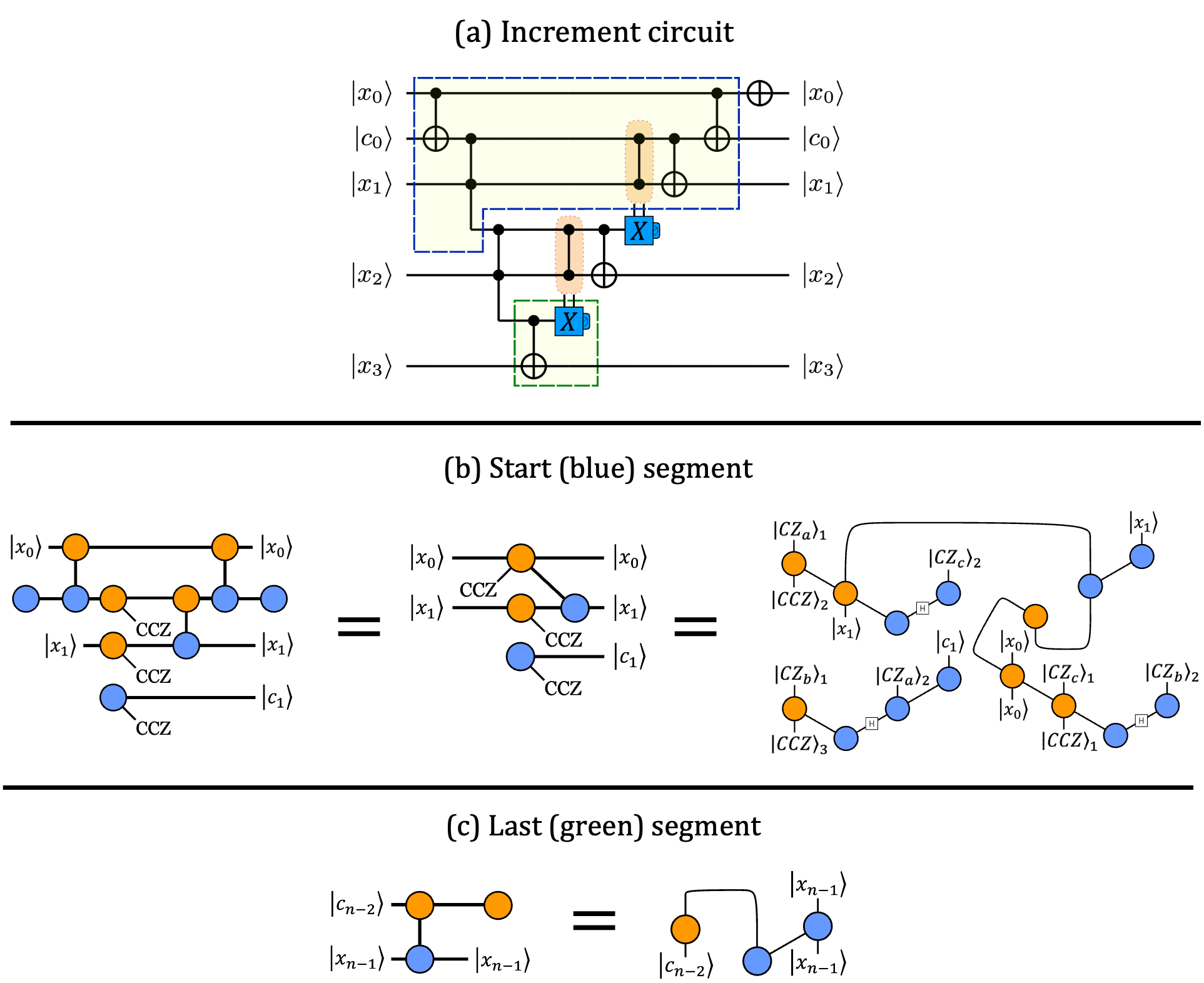}
\caption{(a) The circuit diagram of our increment subroutine. The start and end segments are defined by the blue and green dashed lines, respectively. (b) The active volume calculation for the start segment, it has a cost of $15 + C_{\ket{CCZ}}$ blocks, the same as a repeating segment. (c) The active volume calculation for the end segment, it has a cost of $3$ blocks.}
\label{fig:increm_start_end_cost}
\end{figure}

\subsection{CAS adder}
\label{sec:CAS_adder_cost}
\begin{figure*}[h]
\centering
\includegraphics[scale=0.7]{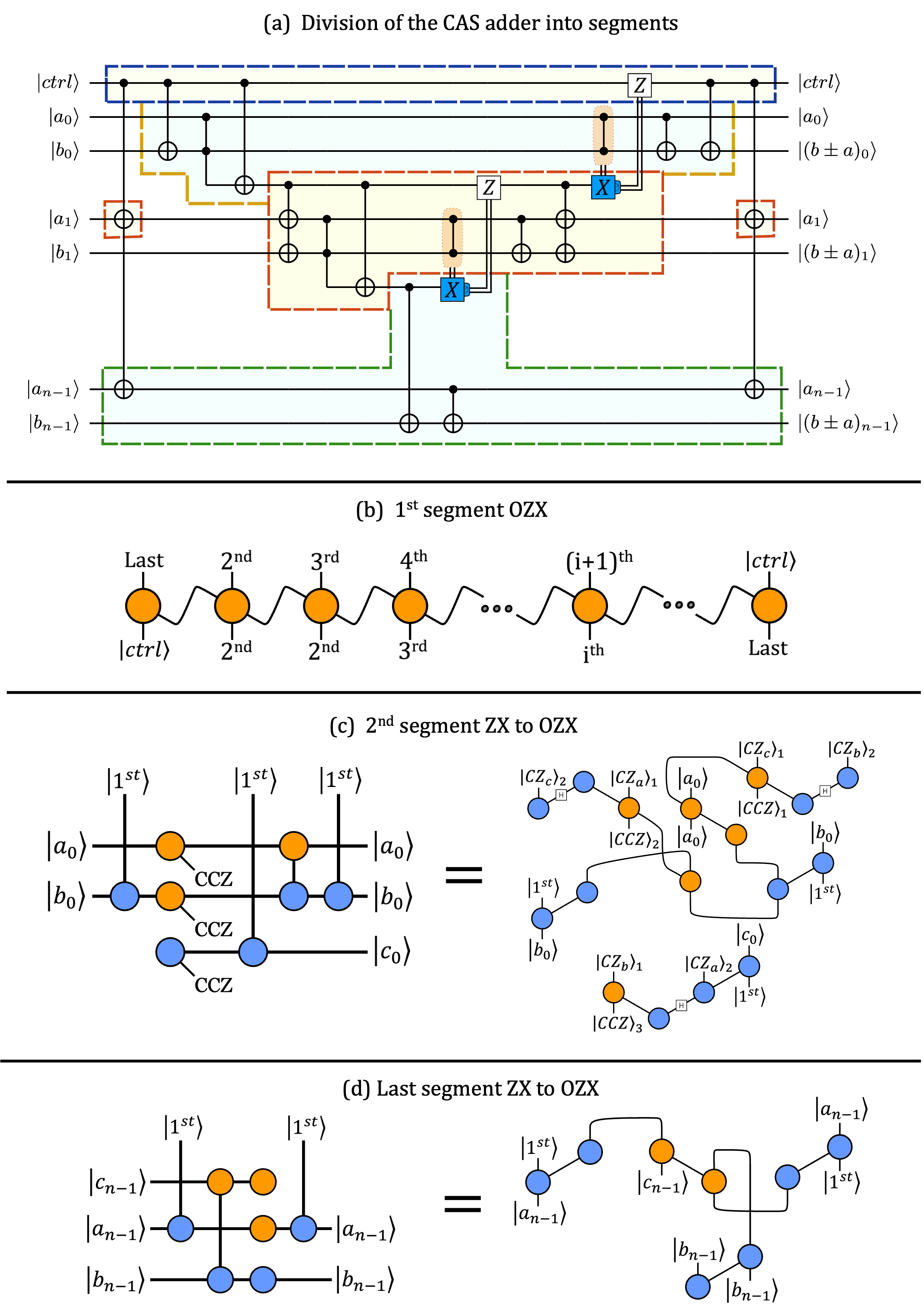}
    \caption{(a) We divide the controlled addition or subtraction (CAS) adder into four different segments, the blue box represents the 1st segment, the orange box represents the 2nd segment, the red boxes represent the repeating segments of which there are $n-2$, and the green box represents the last segment. The repeating segment is identical to the repeating adder segment from \cite{litinski2022active} but has two additional CNOT operations which adds $2$ additional blocks; the block cost is therefore $24 +C_{\ket{CCZ}}$. (b) The first segment contains only controls and therefore conversion to its OZX form is trivial. Each Z spider can hold two connections to X spiders in other segments, apart from the spiders which also have $\ket{ctrl}$ as an input/output, this means the first segment has $n+2$ blocks. (c) The active volume of the second segment is $17+C_{\ket{CCZ}}$ blocks, the T count is $7$, and the reaction depth is $2$. (d) the active volume of the last segment is 9 blocks.}
       \label{fig:CAS_active_volume}
\end{figure*}

Here we calculate the cost of our controlled addition or subtraction subroutine. We can immediately remove 4 CNOT gates from the top two rails of the design displayed in Fig.~\ref{fig:CAS_circuit} as these pairs cancel out. Now we define four different segments, see Fig.~\ref{fig:CAS_active_volume}. \textbf{First segment.} The first segment consists of $2(n-1) + 3$ controls on the top rail and therefore $2(n-1) + 3$ Z spiders. Each of these Z spiders is E-oriented and connects via the U or D port to an N-oriented X spider. The first and last spiders can add $1$ extra connection as they have an input or output, the rest can take $2$ connections to X spiders in other segments, see Fig.~\ref{fig:CAS_active_volume} (a). Adding one connection to the input and output spiders leaves $2(n-1)+1$ connections. It follows that we require an additional $\left\lceil \frac{2n-1}{2} \right\rceil = n$, Z spiders. Therefore, the first segment costs $n+2$ blocks. \textbf{Repeating segments.} The cost of a repeating segment is the same as the cost of the repeating segments from the adder in \cite{litinski2022active}, but with the addition of two more X spiders. Therefore, the block cost of a repeating segment is $24+C_{\ket{CCZ}}$, T count $7$, reaction depth $2$. The total of block cost, T count, and reaction depth of the CAS adder is given by summing the cost of each segment:
\begin{equation}
V_{CAS} = n+2 + 17 + C_{\ket{CCZ}} + (n-2)(24+ C_{\ket{CCZ}}) + 9 = (25+ C_{\ket{CCZ}})n-20- C_{\ket{CCZ}}, \notag
\end{equation}
\begin{equation}
T_{CAS} = 7 + 7(n-2) = 7n-7, \notag
\end{equation}
\begin{equation}
D_{CAS} = 2 + 2(n-2) = 2n-2. \notag
\end{equation}

\begin{figure*}[t]
\centering
\includegraphics[scale=0.8]{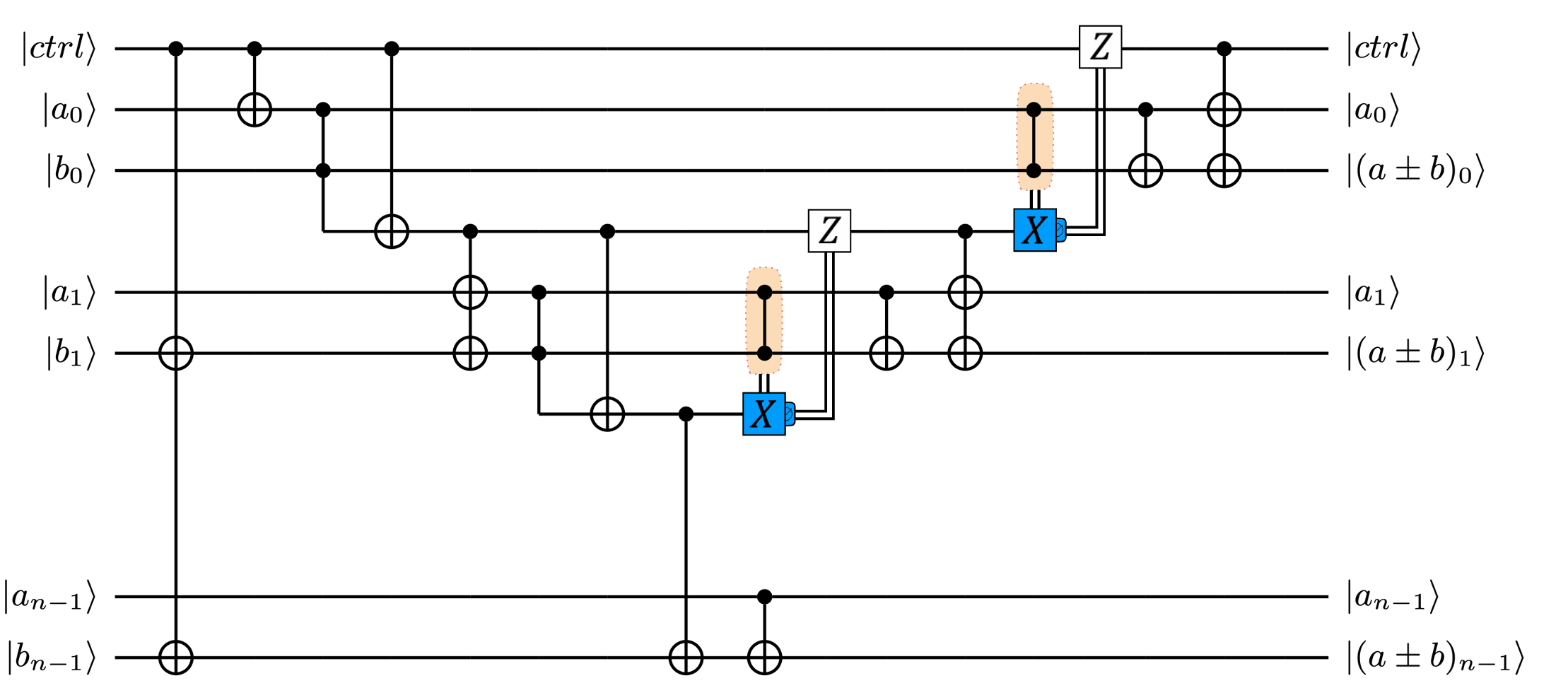}
    \caption{The circuit design for a cheaper CAS adder that performs $a \pm b$ and stores the result in register $\ket{b}$. This version of the circuit requires $n$ fewer CNOT gates.}
       \label{fig:CAS_circuit_cheaper}
\end{figure*}

\subsection{COG adder}
\label{sec:COG_adder_cost}
\begin{figure*}
\centering
\includegraphics[scale=0.95]{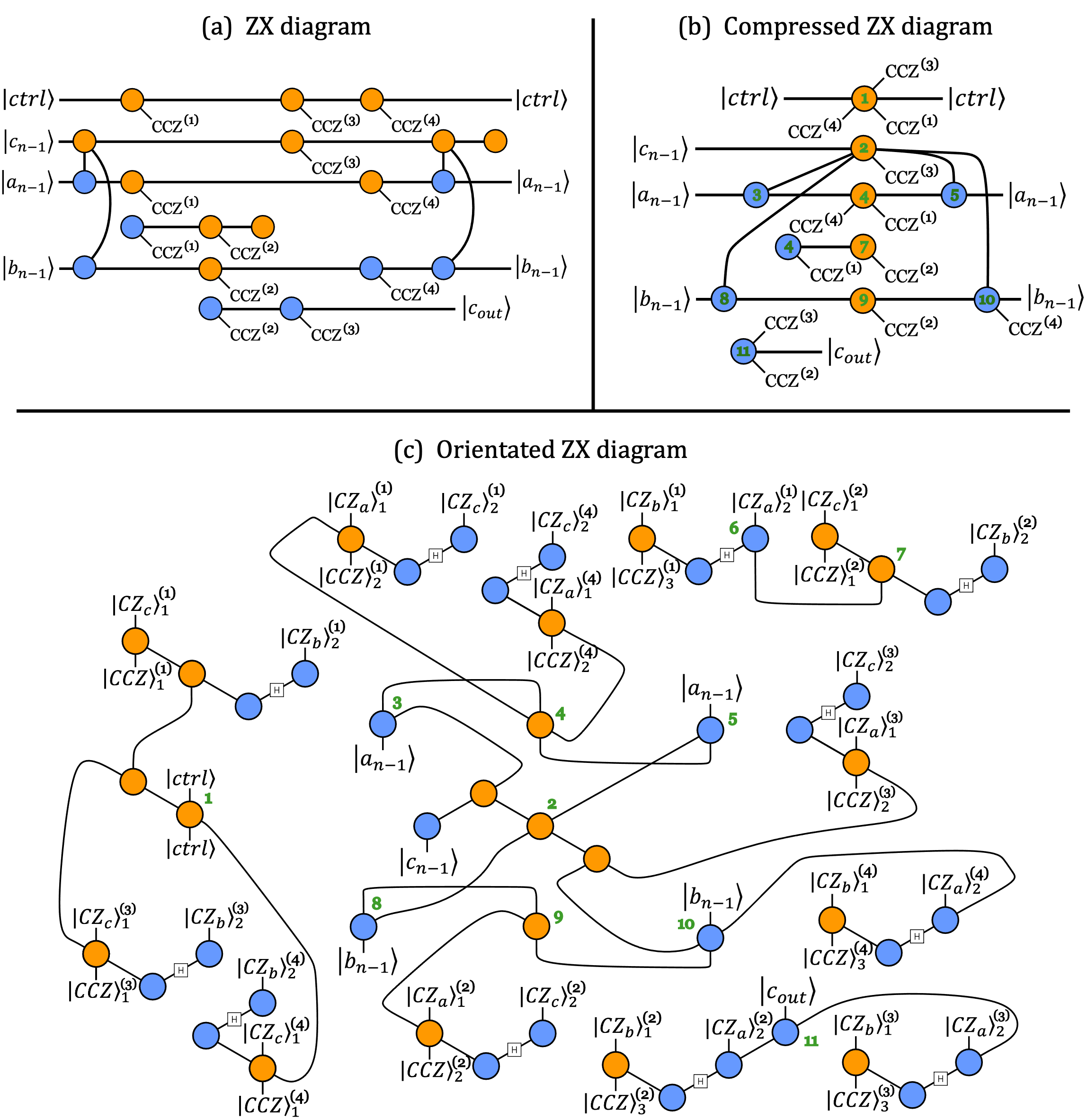}
    \caption{(a) The ZX diagram for the controlled output carry Gidney (COG) adder’s end segment. (b) As the end segment ZX diagram is complex we simplify the things by creating a compressed ZX diagram. To help the construction of the OZX diagram (and to help guide the reader) we number each spider. (c) The orientated ZX diagram for the COG adder end segment. Each number that is attached to an island corresponds to the spider with the same number on the compressed ZX diagram. The cost of the end segment is $51$ blocks and four CCZ states, i.e. $51+4C_{\ket{CCZ}}$. The T count is $28$, and the reaction depth is $4$.}
       \label{fig:COG_end_seg_ZX}
\end{figure*}

Here we calculate the cost of a controlled Gidney adder with output carry. The start and repeating segments are the same as the controlled adder in \cite{litinski2022active} and therefore have a block cost of $24+2C_{\ket{CCZ}}$ and $30+2C_{\ket{CCZ}}$, a T count of $14$ and $14$, and a reaction depth of $2$ and $3$, respectively. 
To determine the cost of our end segment, Fig.~\ref{fig:COG_end_segment}, we must reconstruct the temporary AND (compute) such that we add a third control. This is achieved by adding another temporary AND gate, see Fig.~\ref{fig:k_controlled_NOT}. The active volume calculation is shown in Fig.~\ref{fig:COG_end_seg_ZX}.
In total, there are $n-2$ repeating segments, along with one start segment and one end segment. Therefore, the total block cost, T count, and reaction depth of a COG adder is given by:

\begin{equation}
V_{COGA} = 24+2C_{\ket{CCZ}} + (n-2)(30+2C_{\ket{CCZ}}) + (51+4C_{\ket{CCZ}})= (30+2C_{\ket{CCZ}})n+15+2C_{\ket{CCZ}}, \notag
\end{equation}
\begin{equation}
T_{COGA} = 14+ 14(n-2)+ 28= 14n+14, \notag
\end{equation}
\begin{equation}
D_{COGA} = 2 + 3(n-2) +4 = 3n. \notag
\end{equation}

\subsection{Square root}
\label{sec:sqrt_cost}
\begin{figure*}
\centering
\includegraphics[scale=0.95]{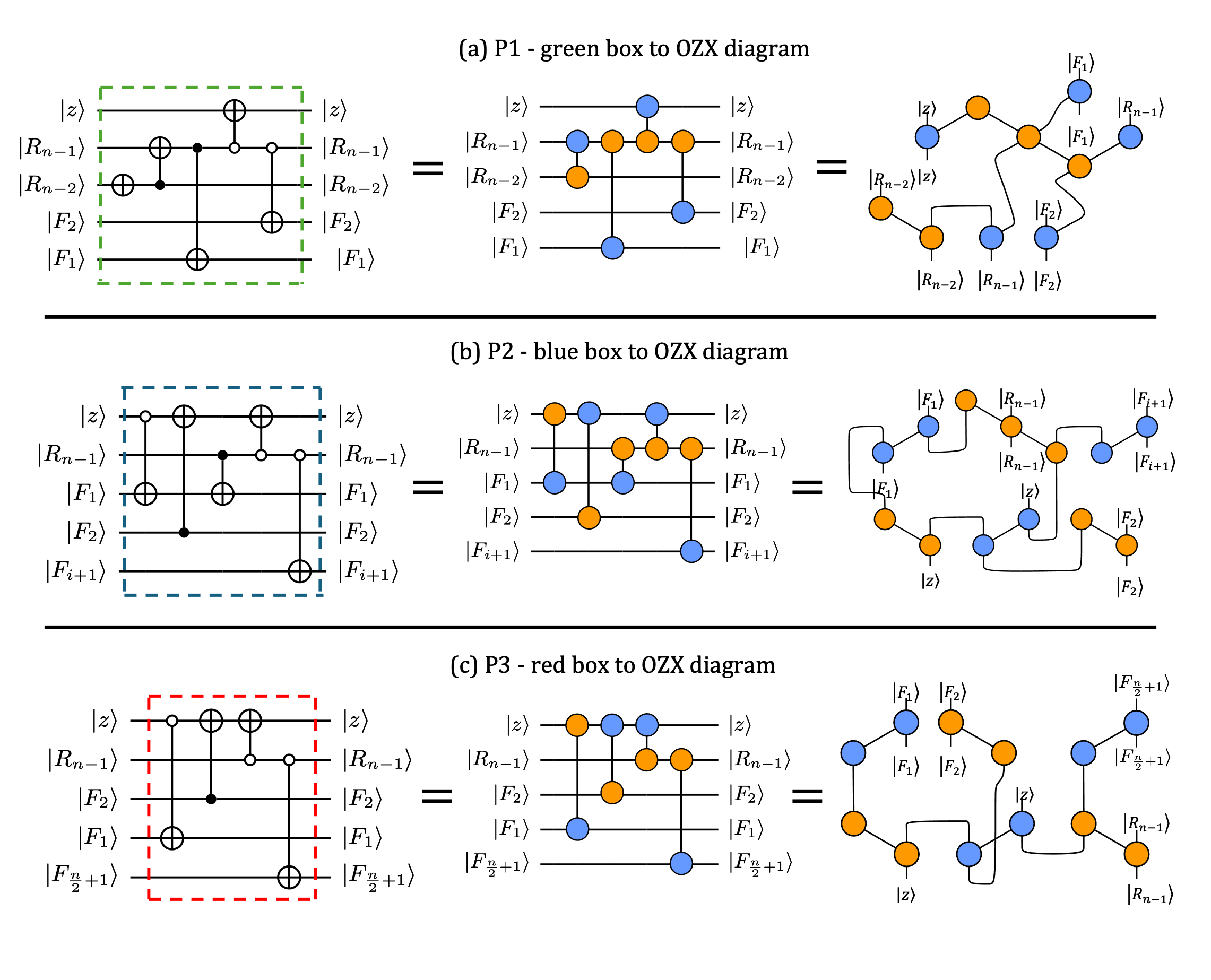}
    \caption{(a) The active volume calculation for the green box in part 1 of the square root circuit shows a cost of $10$ blocks. (b) The active volume calculation for the blue box in part 2 of the square root circuit shows a cost of $13$ blocks. (c) The active volume calculation for the red box in part 3 of the square root circuit shows a cost of $12$ blocks.}
       \label{fig:sqrt_ZX}
\end{figure*}

See Fig.~\ref{fig:sqrt_ZX}.

\subsection{OG adder}
\label{sec:OG_adder_cost}
Here we calculate the cost of a Gidney adder with output carry. The cost of the first segment and repeating segments are $15 + C_{\ket{CCZ}}$ and $22 + C_{\ket{CCZ}}$ blocks, with a T count of $7$ and $4$, and reaction depth $1$ and $2$ \cite{litinski2022active}. The last segment has a block cost of $22 + C_{\ket{CCZ}}$, a T count of $7$, and a reaction depth of $1$. There are $n-2$ repeating segment so the total active volume is given by:
\begin{equation}
V_{OGA} = 15 + C_{\ket{CCZ}} + (n-1) \cdot (22 + C_{\ket{CCZ}}) = (22 + C_{\ket{CCZ}})n -7.
\end{equation}
The T count is given by:
\begin{equation}
T_{OGA} =  7 + 7(n-2) + 7 = 7n.
\end{equation}
The depth is given by: 
\begin{equation}
D_{OGA} = 1+ 2(n-1)  + 1= 2n-2.
\end{equation}

\subsection{Comparator}
\label{sec:cmp_cost}
\begin{figure*}
\centering
\includegraphics[scale=0.9]{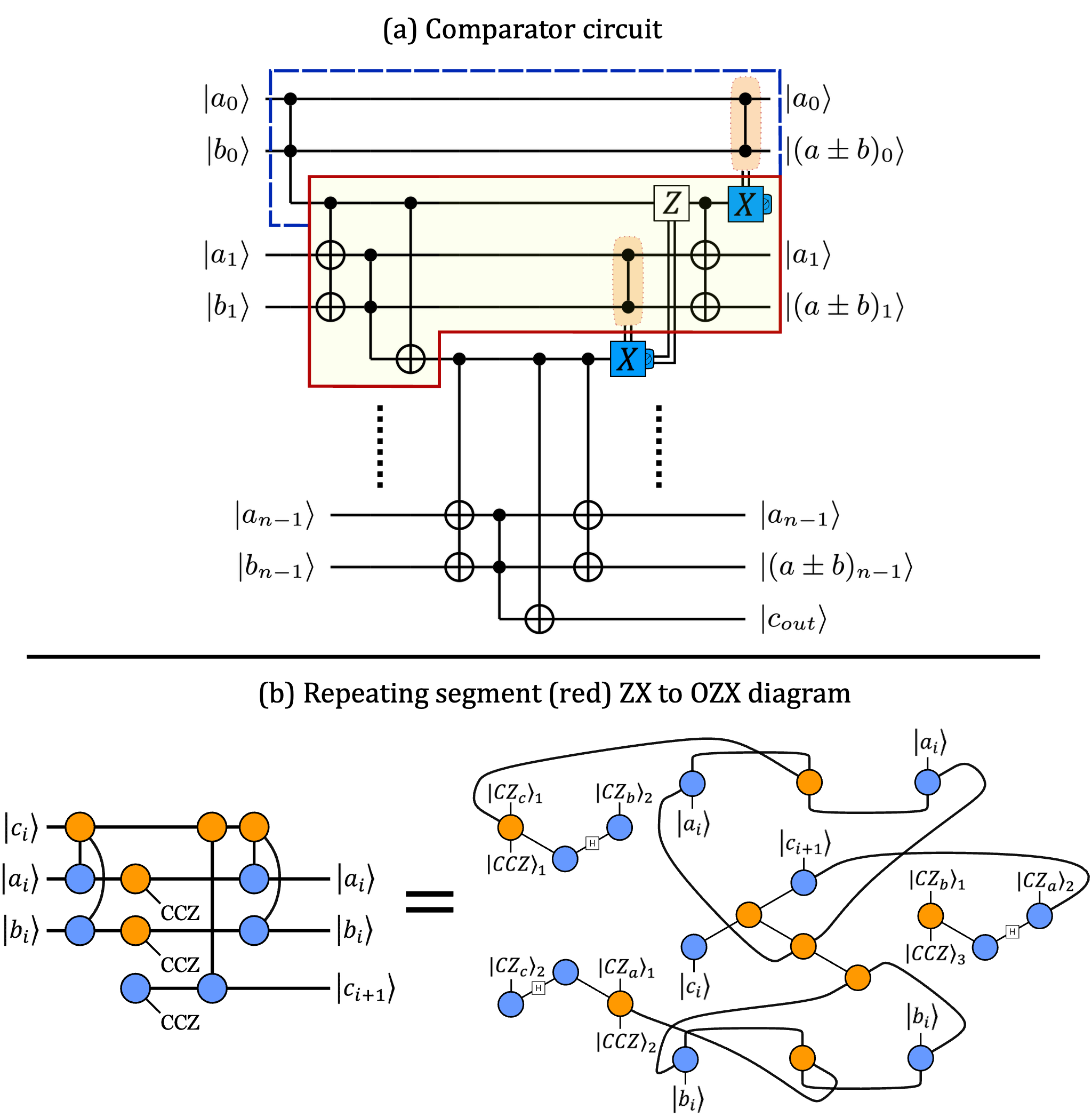}
    \caption{(a) The two component segments of our comparator circuit. The first segment enclosed in the blue dashed line is costs the same as a Toffoli gate. The rest of the circuit is composed of $n-1$ copies of the segment enclosed by the solid red line. (b) The active volume calculation for the red line segment. On the left we represent it as a compressed ZX diagram, on the right we write it as an orientated ZX diagram. The orientated ZX diagram has $20$ spiders and one $\ket{CCZ}$ state as input, so the cost of a repeating segement is $20+C_{\ket{CCZ}}$. The total cost of a comparison is $(20+C_{\ket{CCZ}})n-8$ blocks.}
       \label{fig:cmp_av_derivation}
\end{figure*}

Here we calculate the cost of our comparator circuit. The first segment is just a Toffoli gate and so costs $12 + C_{\ket{CCZ}}$ blocks. The rest of the circuit is made up of $n-1$ repeating segments, where $n$ is the number of qubits in registers $a$ and $b$. The cost of a repeating segment is $20 + C_{\ket{CCZ}}$, see Fig.~\ref{fig:cmp_av_derivation}. The total cost of a comparison is $(20 + C_{\ket{CCZ}})n -8$ blocks, with a T count of $7n$, and reaction depth $2n-1$. The total active volume is given by:
\begin{equation}
V_{OGA} = 12 + C_{\ket{CCZ}} + (20+C_{\ket{CCZ}})(n-1)= (20 + C_{\ket{CCZ}})n -8.
\end{equation}
The T count is given by:
\begin{equation}
T_{OGA} =  7 + 7(n-2) + 7 = 7n.
\end{equation}
The depth is given by: 
\begin{equation}
D_{OGA} = 1 + 2(n-1) = 2n-1.
\end{equation}

\section{Derivations}
\label{sec:derivations}
\subsection{Controlled shift-by-one optimisation}
\label{sec:cshift_control_optimisation}
\begin{figure*}
\centering
\includegraphics[scale=0.95]{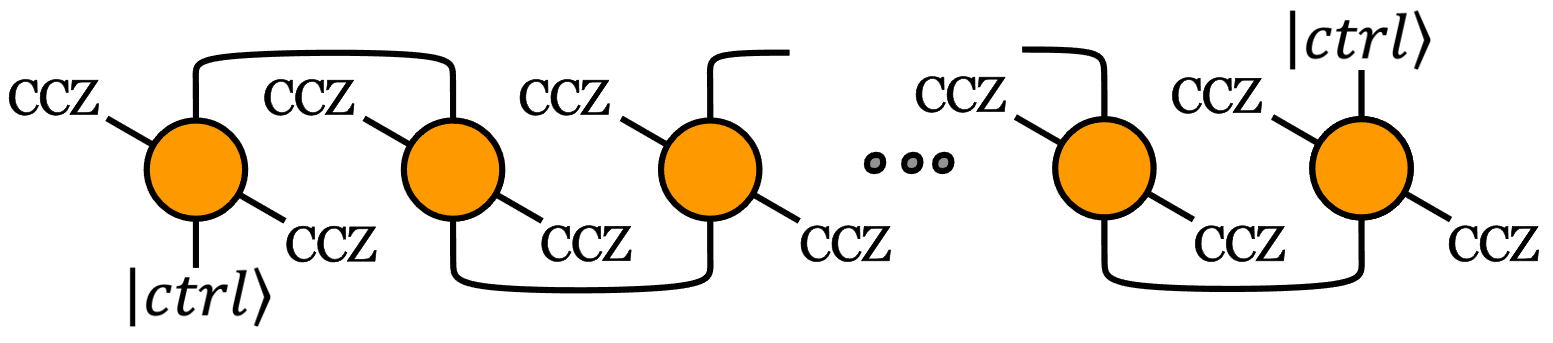}
    \caption{The optimal Z spider configuration for a controlled shift’s control qubit. For an $n$-qubit shift there are $n$ CCZ states. This means the total number of blocks for the control qubit is given by $2 \left\lceil \frac{n-2}{4} \right\rceil +1$.}
       \label{fig:cshift_control_opt}
\end{figure*}

Prior to any optimisation the control qubit rail has $n$ Z spiders each connected via the N or S port to a CCZ island. The optimal configuration is shown in Fig.~\ref{fig:cshift_control_opt}. In this configuration the number of Z spiders must increase by two each time we require more CCZ connections. This is due to the fixed port connections attached to the input and output states of $\ket{a}$. An additional two Z spiders increase the number of N and S ports by four. Therefore, for $n > 2$ we require $2 \left\lceil \frac{n-2}{4} \right\rceil +1$, where the factor of $2$ appears because the number of Z spiders required increases in multiples of $2$, the division by $4$ is because we add $4$ connections when the Z spider count is increased, the $n-2$ and $+1$ is because the repetitive sequence begins at three CCZ island connections. As we originally had $n$ blocks the optimised active volume is given by:
\begin{equation}
V_{cshift} = (20 + C_{\ket{CCZ}})n -(n-2 \left\lceil \frac{n-2}{4} \right\rceil +1).
\end{equation}

\subsection{H\"{a}ner square root calculation}
\label{sec:haner_sqrt_cost_comparison}
Here we calculate the cost reduction given by replacing H\"{a}ner’s square root method (using the cheap multiplication method) with our square root subroutine.
H\"{a}ner computes the square root of $x$ by $\sqrt{x}=x\cdot \frac{1}{\sqrt{x}}$ and uses an adder with Toffoli ($\tau$) count $2n-1$ and a controlled adder with Toffoli count $3n+3$ \cite{takahashi2009adder}. The Toffoli count of a H\"{a}ner multiplication is $\tau_{mult} = 3n^2 +3n$. The Toffoli count of the reciprocal square root, $\frac{1}{\sqrt{x}}$, is given by $\tau_{invsqrt} = n^2(\frac{15}{2}m+3)+15npm+n(\frac{23}{2}m+5)-15p^2 m+15pm-2m$, where $m$ is the number of newton Raphson iterations. We use $m=3$ as this this is the middle value from H\"{a}ner’s simulations \cite{haner2018polyfit}:
\begin{equation}
\tau_{invsqrt} = \frac{51}{2}n^2+\frac{79}{2}n+45np+45p-45p^2 -6. \notag
\end{equation}
Therefore, the Toffoli count of a H\"{a}ner square root is given by:
\begin{equation}
\tau_{sqrt} = \tau_{mult} +\tau_{invsqrt}  = 27n^2+41n+48np+48p-48p^2 -6=D_{sqrt}^{\text{H\"{a}ner}}. \notag
\end{equation}
Finally, we convert this to a T count by multiplying by $7$:
\begin{equation}
T_{sqrt}^{\text{H\"{a}ner}} = 189n^2+287n+336np+336p-336p^2 -42. \notag
\end{equation}
The reduction is the difference in cost compared to our own implementation:
\begin{equation}
T_{reduced} = T_{sqrt}^{\text{H\"{a}ner}} -(1.75n^2 +17.5n+14)= 187.25n^2+269.5n+336np+336p-336p^2 -56, \notag
\end{equation}
\begin{equation}
D_{reduced} = D_{sqrt}^{\text{H\"{a}ner}} -(\frac{1}{2}n^2 +3n-4)= \frac{53}{2}n^2+38n+48np+48p-48p^2 -2. \notag
\end{equation}

\subsection{PPE: the number of CNOT gates for the label circuit}
\label{sec:label_CNOTs}
Consider an $n$-qubit number in binary representation. We wish to increment this number from $0$ to $2^n-1$. Let $a_{n-1}$ be the number of CNOT operations required to fully increment from $0,1,2,\ldots,2^{n-1}-1$. The binary representation of the integer $2^{n-1}-1$: is as follows all qubits are 1 apart from the MSB (e.g. $01\ldots111$).  Incrementing to the next integer, $2^{n-1}$, requires $n$ CNOTs as we use $n-1$ CNOTs flip the $1$’s to $0$’s and one CNOT flips the $n$th qubit to a $1$ (e.g. $10\ldots00$). To finishing incrementing the register to $2^n-1$ we require another $a_{n-1}$ CNOT operations such that each qubit in the register has a value of $1$ (e.g. $11\ldots111$). Therefore, the total number of CNOTs required is 
\begin{equation} 
a_{n}=2a_{n-1}+n.
\label{eq:recur_relation}
\end{equation}
This is a recurrence relation that we can solve. The homogeneous solution is trivial and given by:
\begin{equation} 
a_{n}=2a_{n-1} = C \cdot 2^n. \notag
\end{equation}
For the particular solution we use the ANSATZ $a_{n}=An+B$ and plug this into Eq.~\eqref{eq:recur_relation} to obtain
\begin{equation} 
An +B =2(A(n-1)+B)+n. \notag
\end{equation}
$A$ can be found by equating the coefficients of $n$:
\begin{equation} 
A =2A+1 = -1. \notag
\end{equation}
$B$ can be found by equating the constants:
\begin{equation} 
B =2B-2A=-2. \notag
\end{equation}
The addition of the particular and homogeneous solutions gives us the following general solution 
\begin{equation} 
a_{n} = C \cdot 2^n -n-2. \notag
\end{equation}
Using $a_{0}=1$ we find $C=2$, and so the general solution simplifies to
\begin{equation} 
a_{n} = 2^{n+1} -n-2, \label{eq:Cnot_gen_sol}
\end{equation}
where $a_{n}$ is the total number of CNOTs required to increment a $n$-qubit register from $0,1,2,\ldots,n-1$

\subsubsection{The exact number of CNOTs}
The previous proof found the number of CNOTs required for the label finding circuit, Section~\ref{sec:poly_eval_circuit}, when $M$ is a power of $2$. Here we provide an iterative process to obtain the exact number of CNOTs for any $M>1$. The method requires us to have chosen a value for $M$ and was therefore not used in the main calculation.
A decimal number $M$ requires $\left\lceil \log_2 M \right\rceil$ bits to be displayed in binary. 
\begin{equation}
n = \left\lceil \log_2 M \right\rceil. \notag
\end{equation}
Now consider $k=\left\lceil \log_2 M \right\rceil -1=n-1$, where $M \geq 1$ is an integer. If we fully increment $k$ bits we produce a register filled with $1$’s up to the $k$th qubit (e.g. for $k=3$: $0\ldots0111$). Another increment would require $k+1=n$ CNOTs (e.g. producing $0\ldots1000$). The MSB now has the correct value for displaying $M$ in binary. The total number of CNOTs is given by the sum of Eq.~\eqref{eq:Cnot_gen_sol} and $k+1$:
\begin{equation}
2^{k+1} -k-2+n = 2^{k+1}-1. \notag
\end{equation}
We can apply the same logic to find the number of CNOTs required for the next digit in the binary representation of $M$. This creates an iterative process to which allows us to determine the number of CNOTs by summing the number of CNOTs required to produce each digit starting from the MSB:
\begin{equation}
n_i = \left\lceil \log_2 M^{(i)} \right\rceil, \notag
\end{equation}
\begin{equation}
M_{i+1} = M^{(i)} -2^{n_i}, \notag
\end{equation}
\begin{equation}
N_{CNOT} = 2^{n_{0}+1}-1+\dots+2^{n_{m}+1}-1= \sum_{i=0}^{m} 2^{n_{i}+1}-1,
\end{equation}
where $i \in \{0,1,\ldots,m\}$ and $n_{0}> n_{1}> \dots >n_{m}$. The final iteration is denoted by $m$ and is where $n_{m}=\left\lceil \log_2 M^{(m)} \right\rceil=0$. Note, this expression is only correct when $M$ is even as this means $M^{(0)}$ is odd (to display e.g. 6 numbers we use a label register going from 0 to 5). For odd $M$, we simply remove the final CNOT i.e. $N_{CNOT}=\sum_{0}^{m}(2^{n_{i}+1}-1)-1$.

\end{document}